%% file: circuits.tex
\documentclass[twoside]{article}

\usepackage{amsmath}
\usepackage{amssymb}
\usepackage{amsfonts}
\usepackage{qic,epsfig}

\usepackage[hidelinks]{hyperref}
\usepackage{latexsym}
\usepackage{enumerate}
\usepackage{enumitem}
\usepackage{float}
\usepackage{graphicx}
\usepackage{mathtools}
\usepackage{mathptmx}
\usepackage{mathpazo}
\usepackage{mathrsfs}
\usepackage{ifthen}
\usepackage{color}
\usepackage{tikz}
\usepackage{tikz-3dplot}
\allowdisplaybreaks

\newtheorem{problem}{Problem}

\floatstyle{ruled}
\newfloat{algorithm}{htb}{loa}
\floatname{algorithm}{Algorithm}

\newcommand{\beq}{\begin{equation}}
\newcommand{\eeq}{\end{equation}}
\newcommand{\ba}{\begin{array}}
\newcommand{\ea}{\end{array}}
\newcommand{\bea}{\begin{eqnarray}}
\newcommand{\eea}{\end{eqnarray}}
\newcommand{\beba}{\begin{equation}\begin{array}{lllll}}
\newcommand{\eeea}{\end{array}\end{equation}}
\newcommand{\bc}{\begin{cases}}
\newcommand{\ec}{\end{cases}}
\newcommand{\bpm}{\begin{pmatrix}}
\newcommand{\epm}{\end{pmatrix}}
\newcommand{\ben}{\begin{enumerate}}
\newcommand{\een}{\end{enumerate}}
\newcommand{\bit}{\begin{itemize}}
\newcommand{\eit}{\end{itemize}}

\newcommand{\mrm}{\mathrm}
\newcommand{\mbb}{\mathbb}
\newcommand{\mcal}{\mathcal}
\newcommand{\mbf}{\mathbf}

\newcommand{\tsf}{\textsf}

\newcommand{\vphi}{\varphi}

\newcommand{\bbR}{\mbb{R}}

\let\originalleft\left
\let\originalright\right
\renewcommand{\left}{\mathopen{}\mathclose\bgroup\originalleft}
\renewcommand{\right}{\aftergroup\egroup\originalright}

\newcommand{\lb}{\left(}
\newcommand{\rb}{\right)}

\newcommand{\lbb}{\left[}
\newcommand{\rbb}{\right]}

\newcommand{\intset}[1]{\left \{1, 2, \dots, #1 \right \}}
\newcommand{\seta}[1]{\left \{ #1 \right \}}
\newcommand{\card}[1]{\left | #1 \right |}
\renewcommand{\emptyset}{\O}

\newcommand{\defeq}{\vcentcolon=}

\newcommand{\mmin}[1]{\mrm{min}\left \{ #1 \right \}}

\newcommand{\dsum}{\displaystyle\sum\limits}

\newcommand{\abs}[1]{\left | #1 \right |}
\newcommand{\sgn}[1]{\operatorname{sgn}\lb #1 \rb}

\newcommand{\spn}[1]{\operatorname{span} \seta{ #1 }}

\renewcommand{\ker}[1]{\operatorname{Ker}\lb #1 \rb}

\newcommand{\range}[1]{\operatorname{Range}\lb #1 \rb}

\newcommand{\floor}[1]{\lfloor #1 \rfloor}

\newcommand{\nm}[1]{\left \| #1  \right \|}
\newcommand{\onenm}[1]{\left \| #1  \right \|_1}

\newcommand{{\matle}}{\preccurlyeq}
\newcommand{{\matge}}{\succcurlyeq}

\newcommand{\poly}[1]{\operatorname{poly}\lb #1 \rb}
\newcommand{\bgo}[1]{O \lb #1 \rb}
\newcommand{\omg}[1]{\Omega\lb #1 \rb}
\newcommand{\tht}[1]{\Theta\lb #1 \rb}
\newcommand{\lto}[1]{o \lb #1 \rb}

\renewcommand{\log}[1]{\operatorname{log} \lb #1 \rb}

\newcommand{\ket}[1]{\left | #1\right\rangle}
\newcommand{\bra}[1]{\left\langle #1\right|}
\newcommand{\braket}[2]{\left\langle #1|#2 \right\rangle}
\newcommand{\ketbra}[2]{\left|#1\rangle \langle #2  \right |}

\newcommand{\zo}{\{0,1\}}

\newcommand{\BQP}{\tsf{BQP}}

\newcommand{\vi}{\mbf{i}}
\newcommand{\vf}{\mbf{f}}
\newcommand{\viext}{\mbf{i}_{ext}}

\newcommand{\uwdg}[1]{\mrm{deg} \lb #1\rb}
\newcommand{\wdg}[1]{\widetilde{\mrm{deg}} \lb #1\rb}

\newcommand{\wdgp}[1]{\widetilde{\mrm{deg}'} \lb #1\rb}

\newcommand{\xor}{\oplus}
\newcommand{\parity}{\mrm{PARITY}}

\newcommand{\proj}[1]{{\Pi}\lb #1 \rb}
\newcommand{\refl}[1]{\mrm{Ref}\lb #1 \rb}

\newcommand{\cond}[1]{\kappa \lb #1 \rb}
\newcommand{\fcond}[1]{\kappa_f \lb #1 \rb}

\newcommand{\lap}{L_G}
\newcommand{\nlap}{\overline{L}_G}
\newcommand{\lapp}{L_{G'}}
\newcommand{\nlapp}{\overline{L}_{G'}}
\newcommand{\efr}{\mrm{R}_{\mrm{eff}}}

\begin{document}

\setlength{\textheight}{8.0truein}    

\runninghead{Efficient Quantum Algorithms for Analyzing Large Sparse Electrical Networks}{Guoming Wang}

\normalsize\textlineskip
\thispagestyle{empty}
\setcounter{page}{1}


\alphfootnote

\fpage{1}

\centerline{\bf
EFFICIENT QUANTUM ALGORITHMS FOR ANALYZING LARGE}
\vspace*{0.035truein}
\centerline{\bf SPARSE ELECTRICAL NETWORKS}
\vspace*{0.37truein}
\centerline{\footnotesize GUOMING WANG}
\vspace*{0.015truein}
\centerline{\footnotesize\it Joint Center for Quantum Information and Computer Science,}
\baselineskip=10pt
\centerline{\footnotesize\it University of Maryland, College Park, MD 20742, USA}

\vspace*{0.21truein}

\abstracts{
Analyzing large sparse electrical networks is a fundamental task in physics, electrical engineering and computer science. We propose two classes of quantum algorithms for this task. The first class is based on solving linear systems, and the second class is based on using quantum walks. These algorithms compute various electrical quantities, including voltages, currents, dissipated powers and effective resistances, in time $\operatorname{poly}(d, c, \operatorname{log}(N), 1/\lambda, 1/\epsilon)$, where $N$ is the number of vertices in the network, $d$ is the maximum unweighted degree of the vertices, $c$ is the ratio of largest to smallest edge resistance, $\lambda$ is the spectral gap of the normalized Laplacian of the network, and $\epsilon$ is the accuracy. Furthermore, we show that the polynomial dependence on $1/\lambda$ is necessary. This implies that  our algorithms are optimal up to polynomial factors and cannot be significantly improved. 
}{}{}

\vspace*{10pt}

\keywords{Quantum algorithm, Electrical network, Spectral graph theory, Linear system, Quantum walk}
\vspace*{3pt}

\vspace*{1pt}\textlineskip    

\section{Introduction}
Quantum computers are believed to be more powerful than classical computers, in the sense that quantum algorithms can solve some computational problems exponentially faster than their classical counterparts. So far, this kind of speedup has been mainly demonstrated for two types of problems: simulation of quantum systems (e.g.~\cite{Llo96,AL97,AT03,BACS05,JLP11}), and algebraic or number theoretic problems (e.g.~\cite{Sho94,Sho97,Hal05,Hal07,CD08}). While the first category is quantum in nature, the second category has some group structure so that quantum Fourier transform can be applied to find periodicity. In contrast, many computational problems in natural science and engineering do not possess this kind of structure, and it remains an important task to understand how efficiently quantum algorithms can solve them. 

In this paper, we investigate the power and limitation of quantum algorithms for analyzing (resistive) electrical networks. The problems we consider are as follows. Suppose a connected undirected graph is given such that each edge has associated with it a real positive resistance, and an electric current is injected at some vertices and extracted at some other vertices. The goal is to determine the induced voltage (i.e. potential difference) between two given vertices, or the induced current on a given edge, or the total power disspiated by this graph. We are also interested in computing the \emph{effective resistance} between two given vertices, which is defined as the induced voltage between these vertices when a unit current is injected at one of them and extracted at the other. These problems are fundamental in physics and electrical engineering, and have numerous applications. Remarkably, they play an important role in computer science as well. The idea of viewing a graph as an electrical network turns out to be very fruitful in the design of fast classical algorithms (e.g.~\cite{SS08,ST08,CKM+10,KLP12,Vis12,LRS13,Mad13,MST15}) and analysis of random walks (e.g.~\cite{DS84,CRR+97}). In recent years, electrical network theory has also begun to be used in the design of efficient quantum algorithms (e.g.~\cite{Bel11,LMS11,Bel12,BR12a,BR12b,BCJ+13,IJ15,JK15,CLM16}) and analysis of quantum walks (e.g.~\cite{Bel13,BCJ+13,Mon15,APVW16}). The ability to quickly compute the above electrical quantities is a requisite for these ideas to work. 

Classically, one calculates the electric potentials and currents in an electrical network as follows \footnote{Once these quantities are known, one can infer the dissipated powers and  effective resistances from them.}. Kirchoff's current law stipulates that the sum of the currents entering a vertex equals the sum of the currents leaving it. Ohm's law states that the voltage across a resistor equals the product of the resistance and the current through it. Combining these two facts, one obtain a system of linear equations. Currently, the best way to solve this linear  system is by using Spielman and Teng's algorithm \cite{ST04, ST06}, which takes nearly linear time in the number of edges. Note that this is nearly optimal for this approach, because simply writing down the vector of potentials (or currents) requires linear time in the number of vertices (or edges), which can be time-consuming for  large graphs. Given this limitation, we naturally ask whether quantum algorithms can perform better on this task.

We answer this question affirmatively by giving a series of efficient quantum algorithms for analyzing large sparse electrical networks. These networks might contain (exponentially) many vertices, but each vertex has only a small number of neighbors  (which can be efficiently found). Such networks frequently arise in both physics and computer science contexts. Our algorithms compute various electrical quantities, including voltages, currents, dissipated powers and effective resistances, in time $\poly{d, c, \log{N}, 1/\lambda, 1/\epsilon}$, where $N$ is the number of vertices in the network, $d$ is the maximum unweighted degree of the vertices, $c$ is the ratio of largest to smallest edge resistance, $\lambda$ is the spectral gap of the normalized Laplacian of the network, and $\epsilon$ is the accuracy. In particular, their dependence on $N$ is exponentially better than that of known classical algorithms. 

Our algorithms can be divided into two classes depending on the main techniques used. The first class of algorithms build certain linear systems and extract useful information from their solutions. These systems include the Laplacian system whose solution encodes the electric potentials, and another linear system whose solution roughly encodes the electric currents. To solve these systems most efficiently, we develop variants of a recent quantum linear system algorithm (QLSA) proposed by Childs, Kothari and Somma \cite{CKS15} (which improves the previous algorithms of Harrow, Hassidim and Lloyd \cite{HHL08} and Ambainis \cite{Amb10}). Previous QLSAs yield a quantum state proportional to the solution of a given linear system, and this has caused some controversy over the years. Our variants output a number, and hence do not have this issue. This number can be the norm of the solution, or the norm of one entry of the solution, or the norm of the difference of two entries of the solution, all of which have a natural physical meaning in the context of our linear systems.

Our second class of algorithms take advantage of the graph structure of the problems under consideration, and beat the first class in computing dissipated powers and effective resistances. They are based on a modern use of quantum walks \cite{Amb03,Sze04}, which are a powerful tool for designing fast quantum algorithms. In the early years, quantum walks were mainly used for amplitude amplification in search problems (e.g.~\cite{Amb03,MSS03,Sze04,MNRS06}). But during recent years, their spectral properties became more and more useful for tackling decision problems (e.g.~\cite{Rei09,Rei10,BR12b,Bel13}). Here we follow the second approach. Specifically, we first establish a relationship between the kernel of the signed weighted incidence matrix of a network and the electrical flow in this network. Then we show that a state encoding this flow can be obtained by performing a boosted version of phase estimation on the quantum walk corresponding to this matrix. We also give efficient implementations of this quantum walk. Finally, we demonstrate how to extract useful information from this state by performing appropriate operations on it. 

As mentioned before, all of our algorithms have polynomial dependence on the parameter $1/\lambda$, where $\lambda$ is the spectral gap of the normalized Laplacian of the network (see Section \ref{sec:graphtheorydefinition} for the precise definition). One may wonder whether this dependence is necessary. We show that this is indeed the case. Specifically, we prove that in order to estimate any of the above electrical quantities within a reasonable accuracy, one has to make $\omg{1/\lambda^k}$ queries to the network, for some constant $k>0$. This lower bound implies that our algorithms are optimal up to polynomial factors and cannot be dramatically improved.

The remainder of this paper is organized as follows. In Section \ref{sec:preliminaries}, we provide some requisite background information, and formally state the problems studied in this work. In Section \ref{sec:algorithmlinearsytem}, we describe a class of quantum algorithms for analyzing electrical networks based on solving linear systems. In Section \ref{sec:algorithmsquantumwalk}, we present another class of quantum algorithms for the same problems based on using quantum walks. In Section \ref{sec:lowerbound}, we prove lower bounds on the quantum query complexity of electrical network analysis. Finally, we conclude in Section \ref{sec:discussion} with some comments and future research directions. 

\section{Preliminaries}
\label{sec:preliminaries}
In this section, we provide the necessary background information to understand this paper. In Section \ref{sec:notation}, we introduce the notation used in this paper. In Section \ref{sec:graphtheorydefinition} and Section \ref{sec:electricalflow}, we give some basic results in spectral graph theory and electrical network theory, respectively. In Section \ref{sec:problems}, we formally state the problems studied in this work. 

\subsection{Notation}
\label{sec:notation}
Given a set $U$, we use $\bbR^U$ to denote the set of all functions from $U$ to $\bbR$. If $U$ is finite, we also treat any $f \in \bbR^U$ as a $\card{U}$-dimensional vector in the natural way. 

Given a real number $z$, we define $\sgn{z}=1$ if $z \ge 0$, and $-1$ otherwise. Given two real numbers $a$, $b$ and a real number $\delta>0$, we say that $a$ is a $\delta$-\emph{additive} approximation of $b$ if $\abs{a-b}\le \delta$, and say that $a$ is a $\delta$-\emph{multiplicative} approximation of $b$ if $\abs{a-b} \le \delta \abs{b}$. We say that an algorithm estimates a quantity $x$ up to additive error $\delta$ if it outputs a $\delta$-additive approximation of $x$, and say that an algorithm estimates a quantity $x$ up to multiplicative error $\delta$ if it outputs a $\delta$-multiplicative approximation of $x$.

We will use the Dirac notation to describe both quantum states and abstract vectors. Namely, depending on the context, $\ket{\vphi}$ can be a (possibly unnormalized) state or a vector in a Hilbert space, and $\bra{\vphi}$ is its conjugate transpose. Moreover, if we write $\ket{\psi} \perp \ket{\vphi}$, we mean that $\braket{\psi}{\vphi}=0$.

Given a vector $x$, we use $\nm{x}$ to denote the $l^2$ norm of $x$. Given a matrix $A$, we use $\nm{A}$ to denote the spectral norm of $A$.

Given a matrix $A$, we say that $A$ is $d$-sparse if each row and column of $A$ contains at most $d$ nonzero entries. Moreover, we use $\range{A}$ to denote the range (i.e. column space) of $A$, and use $\ker{A}$ to denote the kernel (i.e. null space) of $A$.  We also use $\proj{A}$ to denote the projection onto $\range{A}$, and use $\refl{A}$ to denote the reflection about $\range{A}$, i.e. $\refl{A} \defeq 2\proj{A}-I$. We also use $s_j(A)$ to denote $j$-th smallest singular value of $A$ (counted with multiplicity), and use $\lambda_j(A)$ to denote the $j$-th smallest eigenvalue of $A$ (counted with multiplicity), starting with $j=1$. The \emph{condition number} of $A$, denoted by $\cond{A}$, is defined as the ratio of largest to smallest singular value of $A$, and the \emph{finite condition number} of $A$, denoted by $\fcond{A}$, is defined as the ratio of largest to smallest nonzero singular value of $A$. Futhermore, we use $A^+$ to denote the \emph{Moore-Penrose pseudoinverse} of $A$. That is, if $A$ has the singular value decomposition $A=\sum_j s_j \ketbra{u_j}{v_j}$ (where $s_j>0$), then $A^+ \defeq \sum_j s_j^{-1} \ketbra{v_j}{u_j}$.

Given two Hermitian matrices $A$ and $B$, if we write $A \matge B$ (or $A \matle B$), we mean that $A-B$ (or $B-A$) is positive semidefinite.

Given a unitary operation $U$ and a real number $\epsilon>0$, we say that a circuit (or procedure) implements $U$ with precision $\epsilon$ if this circuit (or procedure) implements a unitary operation $V$ satisfying $\nm{U-V} \le \epsilon$. 

\subsection{Graph theory definitions}
\label{sec:graphtheorydefinition}
All the graphs considered in this paper will be \emph{connected}, \emph{weighted} and \emph{undirected}, unless otherwise stated. If any unweighted graph is mentioned,  we also treat it as a weighted graph with unit edge weights.

Let $G=(V, E, w)$ be a graph with edge weights $w_e>0$. For any vertex $v$, let $E(v)$ be the set of edges incident to $v$. The \emph{unweighted degree} of $v$ is defined as $\uwdg{v} \defeq \card{E(v)}$, and the \emph{weighted degree} of $v$ is defined as $\wdg{v} \defeq \sum_{e \in E(v)} w_{e}$. The \emph{maximum unweighted degree} of $G$ is defined as $\uwdg{G} \defeq \mrm{max}_{v \in V} \uwdg{v}$, and the \emph{maximum weighted degree} of $G$ is defined as $\wdg{G} \defeq \mrm{max}_{v \in V} \wdg{v}$. For any $S \subseteq V$, the \emph{volume} of $S$ is defined as $\mrm{vol}(S) \defeq \sum_{v \in S} \wdg{v}$. 

Now we arbitrarily orient the edges in $E$. For each edge $e$, let $e^+$  denote its head, and let $e^-$ denote its tail. For each vertex $v$, let $E^+(v) \defeq\{e \in E(v):~e^+=v\}$, and let $E^-(v)\defeq\{e \in E(v):~e^-=v\}$. These orientations are merely for notational convenience, and they are used to interpret the meaning of a positive flow on an edge. That is, if the flow runs from the tail to the head of the edge, then it is positive; otherwise, it is negative. One should keep in mind that the graph $G$ is still undirected, and the flow on an edge can go in either direction, regardless of this edge's orientation.

Now we define several matrices associated with the graph $G$. The \emph{weighted degree matrix} of $G$ is defined as
\beba
D_G \defeq \dsum_{v \in V} \wdg{v} \ketbra{v}{v}.
\eeea
The \emph{weighted adjacency matrix} of $G$ is defined as
\beba
A_G \defeq \dsum_{e \in E} w_e \lb \ketbra{e^-}{e^+} + \ketbra{e^+}{e^-} \rb.
\eeea
The \emph{signed (vertex-edge) incidence matrix} of $G$ is defined as
\beba
B_G \defeq \dsum_{e \in E} \lb \ket{e^-} - \ket{e^+} \rb \bra{e}.
\eeea
The \emph{edge weight matrix} of $G$ is defined as
\beba
W_G \defeq \dsum_{e \in E} w_e \ketbra{e}{e}.
\eeea
The \emph{signed weighted (vertex-edge) incidence matrix} of $G$ is defined as
\beba
C_G  \defeq B_G W_G^{1/2}
= \dsum_{e \in E} \sqrt{w_e}(\ket{e^-} - \ket{e^+})\bra{e}.
\eeea
The \emph{Laplacian} of $G$ is defined as
\beba
\lap \defeq C_G C_G^T= B_GW_GB_G^T = D_G - A_G.
\eeea
The \emph{normalized Laplacian} of $G$ is defined as
\beba
\nlap \defeq D_G^{-1/2}\lap D_G^{-1/2} = I - D_G^{-1/2}A_GD_G^{-1/2}.
\eeea

Both $\lap$ and $\nlap$ are real symmetric matrices, and they satisfy
\bea
&\ker{\lap}=\spn{\ket{\mbf{1}}}, \\
&\ker{\nlap}=\spn{D_G^{1/2}\ket{\mbf{1}}},
\eea
and
\bea
&\range{\lap}=\seta{\ket{\psi}:~\ket{\psi} \perp \ket{\mbf{1}}}, \\
&\range{\nlap}=\seta{D_G^{-1/2}\ket{\psi}:~\ket{\psi} \perp \ket{\mbf{1}}},
\eea
where $\ket{\mbf{1}} \defeq \sum_{v \in V} \ket{v}$. Furthermore, it can be shown that
\beba
0=\lambda_1(\lap) < \lambda_2(\lap)\le \dots \lambda_N(\lap) \le 2 \wdg{G}
\label{eq:eigvallap}
\eeea
and 
\beba
0=\lambda_1(\nlap) < \lambda_2(\nlap) \le \dots \lambda_N(\nlap) \le 2, 
\label{eq:eigvalnlap}
\eeea
where $N \defeq \card{V}$. In particular,  $\lambda_2(\lap)$ is called the \emph{spectral gap} of $\lap$, and $\lambda_2(\nlap)$ is called the \emph{spectral gap} of $\nlap$. 

The following lemma establishes a relationship between $\lambda_2(\lap)$ and
$\lambda_2(\nlap)$:

\begin{lemma}
If $\wdg{v} \ge 1$ for all $v \in V$, then $\lambda_2(\lap) \ge \lambda_2(\nlap)$.
\label{lem:lapnlapspectralgap}
\end{lemma}
\proof{ Suppose $\lambda_2(\nlap)=\lambda$. We need to show that for any $\ket{\psi} \perp \ket{\mbf{1}}$,
 $\braket{\psi}{\psi} \neq 0$, 
\beba
\bra{\psi} \lap \ket{\psi}
\ge \lambda \braket{\psi}{\psi}.
\eeea
 
Since $\ket{\vphi} \defeq D_G^{-1/2}\ket{\psi} \in \range{\nlap}$ and $\lambda_2(\nlap)=\lambda$, we have
\beba
\nlap \matge \lambda \dfrac{\ketbra{\vphi}{\vphi}}{\braket{\vphi}{\vphi}}
= \lambda \dfrac{D_G^{-1/2}\ketbra{\psi}{\psi} D_G^{-1/2}}{\bra{\psi}D_G^{-1}\ket{\psi}}.
\eeea
Consequently, we get
\beba
\bra{\psi} \lap \ket{\psi}
=\bra{\psi} D_G^{1/2} \nlap D_G^{1/2} \ket{\psi}
\ge \lambda \dfrac{\braket{\psi}{\psi}\braket{\psi}{\psi}}{\bra{\psi}D_G^{-1}\ket{\psi}}
\ge \lambda \braket{\psi}{\psi},
\eeea 
where the last step follows from the fact that $D_G=\sum_v \wdg{v} \ketbra{v}{v} \matge I$ and hence $\bra{\psi}D_G^{-1}\ket{\psi} \le \braket{\psi}{\psi}$.
}

Now we define a combinatorial quantity associated with the graph $G$. For any $S$, $T \subset V$, $S \cap T =\emptyset$, let $E(S, T)$ be the set of edges with one endpoint in $S$ and another endpoint in $T$, and let $\mrm{w}(S, T) \defeq \sum_{e \in E(S, T)} w_e$. Then for any $S \subset V$, $S \neq \emptyset$, the \emph{conductance} of $S$ is defined as
\beba
\phi_S \defeq \dfrac{\mrm{w}(S, \bar{S})}{\mmin{\mrm{vol}(S),\mrm{vol}(\bar{S})}},
\eeea
where $\bar{S} \defeq V \setminus S$. Then the \emph{conductance} of $G$ is defined as
\beba
\phi_G \defeq \min\limits_{S \subset V, ~S \neq \emptyset} \phi_S.
\eeea

Remarkably, Cheeger's inequality \cite{Che70,Chu97} establishes a polynomial relationship between the algebraic quantity $\lambda_2(\nlap)$ and the combinatorial quantity $\phi_G$:
\beba
\dfrac{\phi_G^2}{2} \le \lambda_2(\nlap) \le 2 \phi_G.
\label{eq:cheeger}
\eeea

\subsection{Electrical flows}
\label{sec:electricalflow}
In this paper, we treat a graph $G=(V, E, w)$ with edge weights $w_e>0$ as an electrical network with the same topology and edge resistances $r_e \defeq 1/w_e$ (or equivalently, edge conductances $w_e$), and vice versa. So from now on we will interchange the terms ``graph" and ``electrical network", as they refer to  the same thing.

Let $\mbf{i}_{ext} \in \bbR^V$ satisfy $\mbf{i}_{ext} \perp \mbf{1} \defeq (1,1,\dots,1)^T$. Suppose we inject an electric current of value $\mbf{i}_{ext}(v)$ at vertex $v$, for each $v \in V$ (if $\mbf{i}_{ext}<0$, then we extract an electric current of value $-\mbf{i}_{ext}(v)$ at vertex $v$). The condition $\mbf{i}_{ext} \perp \mbf{1}$ ensures that the total amount of injected currents equals the total amount of extracted currents, which is physically reasonable. Let $\mbf{v} \in \bbR^V$ be the induced potentials at the vertices, and let $\mbf{i} \in \bbR^E$ be the induced currents on the edges. By Ohm's law, the current on an edge is equal to the voltage (i.e. potential difference) between its endpoints times its conductance:
\beba
\mathbf{i}=W_GB_G^T\mathbf{v}.
\label{eq:lapcurrent}
\eeea
Meanwhile, by Kirchoff's current law, the sum of the currents leaving a vertex is equal to the amount injected at the vertex:
\beba
B_G \mbf{i}=\vi_{ext}.
\eeea
Combining these two facts, we get
\beq
\vi_{ext}=B_GW_GB_G^T\mbf{v}=\lap \mbf{v}.
\label{eq:eflapsys}
\eeq
Since $\vi_{ext} \in \range{\lap}$, we have
\beq
\mbf{v}=\lap^+ \vi_{ext}.
\label{eq:eflapsyssol}
\eeq
Furthermore, by Joule's first law, the power dissipated by an edge is equal to the square of the current on it times its resistance (or equivalently, the square of the voltage across it times its conductance). So the total power dissipated by the graph $G$ is 
\beba
\mcal{E}(\mbf{i}) \defeq \mbf{i}^T W_G^{-1} \mbf{i}
=\mbf{v}^T \lap \mbf{v}
=\mbf{i}_{ext}^T \lap^+ \mbf{i}_{ext}.
\label{eq:lappower}
\eeea

Often we are interested in the special case where a unit current is injected at a vertex $s$ and extracted at another vertex $t$. Namely, $\mbf{i}_{ext}=\chi_{s,t} \defeq \ket{s}-\ket{t}$. The \emph{effective resistance} between $s$ and $t$, denoted by $\efr(s,t)$, is defined as the induced voltage between $s$ and $t$ in this case. By Joule's first law, $\efr(s,t)$ is also equal to the power dissipated by the graph $G$ in this case. So we have
\beba
\efr(s,t)=\mathbf{v}(s)-\mathbf{v}(t)=\mcal{E}(\vi)
=\chi_{s,t}^T \lap^{+} \chi_{s,t},
\label{eq:effrst}
\eeea
where $\mbf{v}=\lap^+ \chi_{s,t}$ and $\vi=W_GB_G^T\mbf{v}=W_GB_G^T\lap^+ \chi_{s,t}$.

There is an alternative definition for electrical flow which turns out to be very useful. Let $\mbf{i}_{ext} \in \bbR^V$ satisfy $\mbf{i}_{ext} \perp \mbf{1}=(1,1,\dots,1)^T$. We say that $\mbf{f} \in \bbR^E$ is a flow \emph{consistent} with $\mbf{i}_{ext}$ if it obeys the flow-conservation constraints:
\beba
\dsum_{e \in E^-(v)} \mbf{f}(e)
-\dsum_{e \in E^+(v)} \mbf{f}(e)
=\mbf{i}_{ext}(v), &\forall~v \in V.
\eeea
The \emph{power} of the flow $\mbf{f}$ (with respect to the edge resistances $r_e=1/w_e$) is defined as 
\bea
\mcal{E}(\mbf{f}) \defeq \sum_{e \in E} r_e \mbf{f}^2(e) = \sum_{e \in E} \dfrac{\mbf{f}^2(e)}{w_e}.
\eea
Then, among all the flows consistent with $\mbf{i}_{ext}$, the electrical flow $\mbf{i}$ induced by $\viext$ is the unique flow that minimizes this power function:

\begin{lemma}
Let $\mbf{i}=W_GB_G^T\lap^+ \mbf{i}_{ext}$ be the electrical flow induced by $\mbf{i}_{ext}$. Then any flow $\mbf{f} \in \bbR^E$ consistent with $\mbf{i}_{ext}$ satisfies $\mcal{E}(\mbf{f}) \ge \mcal{E}(\mbf{i})$.
\label{lem:minpower}
\end{lemma}
\proof{ Suppose $C_G$ has the singular value decomposition $C_G=\sum_j s_j \ketbra{u_j}{v_j}$, where $s_j>0$, $\ket{u_j}$ and $\ket{v_j}$ are real vectors, for all $j$. Then $\lap = C_G C_G^T$ has the spectral decomposition 
$\lap=\sum_j s_j^2 \ketbra{u_j}{u_j}$. In addition, since $\mbf{i}_{ext} \in \range{\lap}$, we have $\mbf{i}_{ext}=\sum_j \alpha_j \ket{u_j}$ for some numbers $\alpha_j$'s. Consequently, we get
\beba
W_G^{-1/2}\mbf{i}=W_G^{1/2}B_G^T\lap^+ \mbf{i}_{ext}
=C_G^T \lap^+\mbf{i}_{ext}
=\dsum_j s_j^{-1} \alpha_j \ket{v_j}.
\eeea
So the power of the electrical flow $\mbf{i}$ is 
\beba
\mcal{E}(\mbf{i}) = \nm{W_G^{-1/2} \mbf{i}}^2 = \dsum_j \abs{s_j^{-1} \alpha_j}^2.
\eeea
Meanwhile, for any flow $\mbf{f}$ consistent with $\mbf{i}_{ext}$, we have $B_G \mbf{f} = \mbf{i}_{ext}$ and hence $C_G (W_G^{-1/2} \mbf{f}) = \mbf{i}_{ext}$. This implies that
\beba
W_G^{-1/2} \mbf{f} = \dsum_j s_j^{-1} \alpha_j \ket{v_j} + \ket{\Phi^{\perp}},
\eeea
where $\ket{\Phi^{\perp}}$ is an unnormalized vector satisfying
$\ket{\Phi^{\perp}} \perp \ket{v_j}$ for all $j$. As a result, we obtain
\beba
\mcal{E}(\mbf{f})=\nm{W^{-1/2} \mbf{f}}^2 = \dsum_j \abs{s_j^{-1} \alpha_j}^2 + \nm{\ket{\Phi^{\perp}}}^2 \ge \mcal{E}(\mbf{i}).
\eeea
}

Lemma \ref{lem:minpower} implies that the effective resistance $\efr(s, t)$ between $s$ and $t$ is equal to the minimum power of a flow consistent with $\mbf{i}_{ext}=\chi_{s, t}$. This is an alternative definition of effective resistance.

\subsection{Problem statement} 
\label{sec:problems}
Given an electrical network $G=(V, E, w)$ driven by an external current $\mbf{i}_{ext}$, we are interested in the quantum complexity of the following problems:
\bit
\item Compute the voltage between two vertices $s$ and $t$.
\item Compute the current on an edge $e$.
\item Compute the power dissipated by the graph $G$. 
\item Compute the effective resistance between two vertices $s$ and $t$.
\eit

We will mainly focus on large sparse graphs. Namely, $G$ might contain (exponentially) many vertices, but each vertex has only a small number of neighbors (which can be efficiently found). Our model is as follows. Suppose $V=\{v_1,v_2,\dots,v_N\}$, $E=\{e_1,e_2,\dots,e_M\}$ and $\uwdg{G}=d$. Then we assume there exists a procedure $\mcal{P}_v$ that, on input $(i, k) \in \intset{N} \times \intset{d}$, outputs (the index of) the $k$-th edge incident to $v_i$. We also assume there exists a procedure $\mcal{P}_e$ that, on input $j \in \intset{M}$, outputs (the index of) the two endpoints of $e_j$ as well as the weight of $e_j$. Furthermore, except for computing effective resistances, we assume there exists a produre $\mcal{P}_i$ that prepares the state $\frac{\ket{\viext}}{\nm{\ket{\viext}}}$, where $\ket{\viext} \defeq \sum_{v \in V} \viext(v) \ket{v}$. We assume that $\mcal{P}_v$, $\mcal{P}_e$ and $\mcal{P}_i$ are all efficient, in the sense that they can be implemented in time $\poly{\log{N}}$. 

Formally, we define our Electrical Network Analysis (ENA) problems as follows:

\begin{problem}[ENA-V]
Let $G=(V, E, w)$ be an electrical network such that $\card{V}=N$, $\uwdg{G} \le d$, $1 \le w_e \le c$, for all $e \in E$, and $\lambda_2(\nlap) \ge \lambda >0$. Suppose $G$ is driven by an external current $\mbf{i}_{ext} \in \bbR^{V}$ satisfying $\mbf{i}_{ext} \perp \mbf{1}$ and $\nm{\mbf{i}_{ext}} = 1$  \footnote{For the readers who have skipped Section \ref{sec:notation}, 
$\nm{\mbf{i}_{ext}}$ means the $l^2$ norm of $\mbf{i}_{ext}$.}.
Let $\mbf{v}=\lap^+ \mbf{i}_{ext}$ be the induced potentials at the vertices, and let $\mbf{i}=W_GB_G^T \mbf{v}$ be the induced currents on the edges. Let $\epsilon \in (0, 1)$. Given $s, t \in V$ and access to the procedures $\mcal{P}_v$, $\mcal{P}_e$ and $\mcal{P}_i$, the goal is to estimate $\abs{\mbf{v}(s)-\mbf{v}(t)}$ up to additive error $\epsilon$, succeeding with probability at least $2/3$.
\end{problem}

\begin{problem}[ENA-C]
The assumption is the same as in ENA-V. Given $e \in E$ and access to the procedures $\mcal{P}_v$, $\mcal{P}_e$ and $\mcal{P}_i$, the goal is to estimate $\abs{\mbf{i}(e)}$ up to additive error $\epsilon$, succeeding with probability at least $2/3$.
\end{problem}

\begin{problem}[ENA-P]
The assumption is the same as in ENA-V. Given access to the procedures $\mcal{P}_v$, $\mcal{P}_e$ and $\mcal{P}_i$, the goal is to estimate $\mcal{E}(\mbf{i})$ up to multiplicative error $\epsilon$, succeeding with probability at least $2/3$.
\end{problem}

\begin{problem}[ENA-ER]
The assumption is almost the same as in ENA-V, except that we do not need $\mbf{i}_{ext}$ or $\mcal{P}_{i}$. Given $s, t \in V$ and access to the procedures $\mcal{P}_v$ and $\mcal{P}_e$, the goal is to estimate $\efr(s, t)$ up to multiplicative error $\epsilon$, succeeding with probability at least $2/3$.
\end{problem}

These problems are not completely independent of each other. For example, when $s$ and $t$ are adjacent vertices, we can infer the voltage between $s$ and $t$ from the current on $(s, t)$ by using Ohm's law, and vice versa. So ENA-V and ENA-C are equivalent (up to a query of the weight of $(s, t)$) in this case. Moreover, when $\mbf{i}_{ext}=\chi_{s, t}$, the power of the flow $\mbf{i}$ is equal to the effective resistance $\efr(s,t)$ between $s$ and $t$. So ENA-ER can be viewed as a special case of ENA-P. Despite such connections among these problems, one can see that no two of them are completely equivalent.

Although in the above problems we assume that the edge conductances are in the range $[1,c]$ and the external current $\viext$ has unit $l^2$ norm, this is without loss of generality. Suppose instead that the edge conductances are in the range $[a, ac]$, and $\nm{\viext}=b$, for some constants $a, b>0$. Namely, the edge conductances are rescaled by a factor of $a$, and the external current is rescaled by a factor of $b$. By Eqs.~(\ref{eq:lapcurrent}), (\ref{eq:eflapsyssol}), (\ref{eq:lappower}) and (\ref{eq:effrst}), this would rescale the voltages, currents, disspated powers and effective resistances in the electrical network by a factor of $b/a$, $b$, $b^2/a$ and $1/a$, respectively. So we only need to solve the problems ENA-V, ENA-C, ENA-P and ENA-ER described above, and then multiply their solutions by these factors, resepectively.

We will develop quantum algorithms for solving the above ENA problems. We quantify the resource requirements of these algorithms using two measures. The \emph{query complexity} is the number of uses of the procedures $\mcal{P}_v$, $\mcal{P}_e$ and $\mcal{P}_i$ in the algorithm. The \emph{gate complexity} is the number of 2-qubit gates used in the algorithm. An algorithm is \emph{gate-efficient} if its gate complexity is larger than its query complexity only by a logarithmic factor. Formally, an algorithm with query complexity $Q$ is gate-efficient if its gate complexity is $\bgo{Q \cdot \poly{\log{QN}}}$, where $N=|V|$ is the number of vertices in $G$. All the algorithms presented in this paper will be gate-efficient.

\section{Analyzing Electrical Networks by Solving Linear Systems}
\label{sec:algorithmlinearsytem}

In this section, we describe a class of quantum algorithms for analyzing electrical networks based on solving certain linear systems. These systems include the Laplacian system whose solution encodes the electric potentials, and another linear system whose solution roughly encodes the electric currents. To solve these systems most efficiently, we first develop several variants of a recent quantum linear system algorithm in Section \ref{sec:qlsa}. Then we show how to use them to solve the ENA problems in Section \ref{sec:aenbyqlsa}.

\subsection{Quantum linear system algorithms}
\label{sec:qlsa}
Recently, Childs, Kothari and Somma (CKS) \cite{CKS15} proposed a quantum linear system algorithm (QLSA) which improves the previous algorithms of Harrow, Hassidim and Lloyd (HHL) \cite{HHL08} and Ambainis \cite{Amb10}. Their main result can be summarized as follows:

\begin{theorem}[\cite{CKS15}]
Let $A$ be a $d$-sparse $N \times N$ Hermitian matrix such that all the eigenvalues of $A$ are in the range $D_{\kappa} \defeq [-1, -1/\kappa] \cup [1/\kappa, 1]$. Assume there exists a procedure $\mcal{P}_A$ that runs in time $\poly{\log{N}}$ and on input $(i, j) \in \intset{N} \times \intset{d}$, outputs the location and value of the $j$-th nonzero entry in the $i$-th row of $A$. Let $\vec{b}=(b_1,b_2,\dots,b_N)^T$ be an $N$-dimensional vector. Assume there exists a procedure $\mcal{P}_b$ that runs in time $\poly{\log{N}}$ and produces the state $\ket{\bar{b}} \defeq \frac{\sum_j b_j \ket{j}}{\nm{\sum_j b_j \ket{j}}}$. Let $\vec{x} =(x_1,x_2,\dots,x_N)^T \defeq A^{-1} \vec{b}$, and let $\ket{\bar{x}} \defeq \frac{\sum_j x_j \ket{j}}{\nm{\sum_j x_j \ket{j}}}$. Let $\epsilon \in (0, 1)$. Then there exists a gate-efficient \footnote{In this subsection, we define \emph{gate-efficient} algorithms as follows: An algorithm with query complexity $Q$ is gate-efficient if its gate complexity is $\bgo{Q \cdot \poly{\log{QN}}}$, where $N$ is the dimension of matrix $A$.} quantum algorithm that makes $$\bgo{d \kappa \cdot \poly{\log{\dfrac{d \kappa}{\epsilon}}}}$$ uses of $\mcal{P}_A$ and $\mcal{P}_b$, and
produces a state $\epsilon$-close to $\ket{\bar{x}}$ in $l^2$ norm, succeeding with $\omg{1}$ probability, with a flag indicating success. 
\label{thm:qlssol}
\end{theorem}

CKS mainly focused on how to prepare a state proportional to the solution of a given linear system. But for electrical network analysis, the following problems are actually more relevant: (1) Compute the norm of this solution; (2) Compute the norm of an entry of this solution; (3) Compute the norm of the difference of two entries of this solution. So we develop variants of their algorithm for solving these problems \footnote{CKS actually gave two quantum algorithms that meet the constraints of Theorem \ref{thm:qlssol}, one based on the Fourier approach, and another based on the Chebyshev approach. Our variants are based on the former one. Moreover, after completing this work, we realize that Ref.~\cite{MP15} gave an alternative proof of Lemma \ref{lem:qlssolnom} based on a modification of HHL's algorithm.}:

\begin{lemma}
Under the same assumption as in Theorem \ref{thm:qlssol}, supposing $||\vec{b}||=q$ is known, there exists a gate-efficient quantum algorithm that makes $$\bgo{\dfrac{d \kappa^2}{\epsilon} \cdot \poly{\log{\dfrac{d \kappa}{\epsilon}}}}$$ uses of $\mcal{P}_A$ and $\mcal{P}_b$, and outputs an $\epsilon$-multiplicative approximation of $\nm{\vec{x}}$ with probability at least $2/3$.
\label{lem:qlssolnom}
\end{lemma}

\begin{lemma}
Under the same assumption as in Theorem \ref{thm:qlssol}, supposing $||\vec{b}||=q$ is known, there exists a gate-efficient quantum algorithm that makes $$\bgo{\dfrac{d q^2\kappa^3}{\epsilon^2} \cdot \poly{\log{\dfrac{d q \kappa}{\epsilon}}}}$$ uses of $\mcal{P}_A$ and $\mcal{P}_b$, and outputs an 
$\epsilon$-additive approximation of $\abs{x_i}$, for any given $i \in \intset{N}$, with probability at least $2/3$.
\label{lem:qlssolentry}
\end{lemma}

\begin{lemma}
Under the same assumption as in Theorem \ref{thm:qlssol}, supposing $||\vec{b}||=q$ is known, there exists a gate-efficient quantum algorithm that makes $$\bgo{\dfrac{d q^2\kappa^3}{\epsilon^2} \cdot \poly{\log{\dfrac{d q \kappa}{\epsilon}}}}$$ uses of $\mcal{P}_A$ and $\mcal{P}_b$, and outputs an
$\epsilon$-additive approximation of $\abs{x_i-x_j}$, for any given $i, j \in \intset{N}$, with probability at least $2/3$.
\label{lem:qlssolentrydiff}
\end{lemma}

Before proving these lemmas, let us briefly review the algorithm in Theorem \ref{thm:qlssol}. Then we show how to modify this algorithm to solve the problems in Lemmas \ref{lem:qlssolnom}, \ref{lem:qlssolentry} and \ref{lem:qlssolentrydiff}.

This algorithm uses the following technique to implement a linear combination of unitary operations. Let $M = \sum_j \alpha_j U_j$ be a linear combination of unitary operators $U_j$ with $\alpha_j > 0$ for all $j$. Let $V$ be any unitary operator that satisfies $V \ket{0^m} = \frac{1}{\sqrt{\alpha}} \sum_j \sqrt{\alpha_j } \ket{j}$, where $m$ is a positive integer, 
$\alpha \defeq \onenm{\vec{\alpha}} = \sum_j \alpha_j$. Let $U \defeq \sum_j \ketbra{j}{j}\otimes U_j$. Then $W \defeq V^{\dagger} U V$ satisfies
\bea
W \ket{0^m}\ket{\vphi} &=& \dfrac{1}{\alpha} \ket{0^m} M \ket{\vphi} + \ket{\Phi^{\perp}} \\
&=& \lb \dfrac{\nm{M \ket{\vphi}}}{\alpha} \rb \ket{0^m} \dfrac{M \ket{\vphi}}{\nm{M \ket{\vphi}}} + \ket{\Phi^{\perp}},
\label{eq:lcu}
\eea
where $\ket{\Phi^{\perp}}$ is an unnormalized state (depending on $\ket{\vphi}$) satisfying $\lb \ketbra{0^m}{0^m} \otimes I\rb \ket{\Phi^{\perp}}=0$, for all state $\ket{\vphi}$. Then, if we measure the first $m$ qubits of this state in the standard basis, then with probability $\frac{\nm{M \ket{\vphi}}^2}{\alpha^2}$, the outcome is $0^m$ and we obtain the state $\frac{M \ket{\vphi}}{\nm{M \ket{\vphi}}}$. 

To apply this technique to implement the operator $A^{-1}$, Ref.~\cite{CKS15} finds certain $\alpha_j$'s and $U_j$'s such that $A^{-1} \approx \sum_j \alpha_j U_j$ and each $U_j$ is of the form $e^{-i A t_j}$ for some $t_j \in \bbR$. Specifically, let $\gamma>0$ be arbitrary, and let the function $h(x)$ be defined as
\beba
h(x) \defeq \dsum_{j=0}^{J-1} \dsum_{k=-K}^{K} \alpha(j, k) e^{-i x \beta(j, k)},
\eeea
where
\bea
&\alpha(j, k) \defeq \dfrac{i}{\sqrt{2\pi}} k \delta_y \delta_z^2  e^{-k^2 \delta_z^2/2}, \\ 
&\beta(j, k) \defeq j k \delta_y \delta_z,
\eea
for some $J=\tht{(\kappa/\gamma) \cdot \log{\kappa/\gamma}}$, $K=\tht{\kappa \cdot \log{\kappa/\gamma}}$, $\delta_y=\tht{\gamma/\sqrt{\log{\kappa/\gamma}}}$ and $\delta_z=\tht{1/(\kappa \sqrt{\log{\kappa/\gamma}})}$. Then $h(x)$ is $\gamma$-close to $1/x$ on the domain $D_{\kappa}$, i.e. $\abs{h(x) - x^{-1}} \le \gamma$ for all $x \in D_{\kappa}$. Then since $A$ is a Hermitian matrix with eigenvalues in the range $D_{\kappa}$,
\beba
\nm{h(A)-A^{-1}}= \nm{\dsum_{j=0}^{J-1} \dsum_{k=-K}^{K} \alpha(j, k) e^{-i A \beta(j, k)} - A^{-1}} \le \gamma.
\eeea
It follows that
\beba
\nm{h(A)\ket{\bar{b}} - A^{-1} \ket{\bar{b}}} = \bgo{\gamma}.
\label{eq:fourierapprox1}
\eeea 
Then, since $\nm{A^{-1} \ket{\bar{b}}} \ge 1$, by Lemma \ref{lem:statedistance} in Appendix A, we get
\beba
\nm{\dfrac{h(A) \ket{\bar{b}}}{\nm{h(A) \ket{\bar{b}}}} - \dfrac{A^{-1}\ket{\bar{b}}}{\nm{A^{-1}\ket{\bar{b}}}}} = \bgo{\gamma}.
\label{eq:fourierapprox2}
\eeea
Furthermore, we have
\beba
\alpha \defeq \dsum_{j=0}^{J-1} \dsum_{k=-K}^{K} \abs{\alpha(j, k)} = \tht{\kappa \sqrt{\log{\kappa / \gamma}}},
\eeea
and 
\beba
\abs{\beta(j, k)} \le JK\delta_y \delta_z = \tht{\kappa\cdot \log{\kappa/\gamma}}
\eeea
for all $j$, $k$.

Now we pick $\gamma = \tht{\epsilon}$, and define the operators $V$ and $U$ corresponding to this Fourier approximation of $x^{-1}$. Let $V$ be a unitary operator such that
\beba
V \ket{0^m} = \dfrac{1}{\sqrt{\alpha}} \dsum_{j=0}^{J-1}\dsum_{k=-K}^{K} 
\sqrt{\abs{\alpha(j, k)}} \ket{j, k},
\eeea
where $m=\bgo{\log{JK}}=\bgo{\log{\kappa/\epsilon}}$. Let $U$ be defined as
\beba
U \defeq i \dsum_{j=0}^{J-1} \dsum_{k=-K}^{K} \ketbra{j, k}{j, k} \otimes \sgn{k} e^{-i A \beta(j, k)}.
\eeea 
Ref.~\cite{CKS15} shows that $V$ can be implemented with $\bgo{\kappa \cdot  \poly{ \log{\kappa/\epsilon}}}$ 2-qubit gates, and $U$ can be implemented with precision $\epsilon' (\le \epsilon)$ by a gate-efficient procedure that makes $O(d\kappa \cdot \poly{\log{d \kappa/\epsilon'}})$ uses of $\mcal{P}_A$.

Now we define $W=V^{\dagger} U V$, and get
\beba
W \ket{0^m} \ket{\bar{b}}
&=& \lb \dfrac{\nm{h(A) \ket{\bar{b}}}}{\alpha} \rb \ket{0^m} \dfrac{h(A) \ket{\bar{b}}}{\nm{h(A)\ket{\bar{b}}}} + \ket{\Phi^{\perp}},
\label{eq:lcu2}
\eeea
where $\ket{\Phi^{\perp}}$ is an unnormalized state satisfying 
$\lb \ketbra{0^m}{0^m} \otimes I\rb \ket{\Phi^{\perp}}=0$. 
Then, if we measure the first $m$ qubits of this state in the standard basis, then with probability 
\beba
p \defeq \dfrac{\nm{h(A)\ket{\bar{b}}}^2}{\alpha^2} = \omg{\dfrac{1}{\alpha^2}},
\eeea
(note that $\nm{h(A) \ket{\bar{b}}} \ge \nm{A^{-1}\ket{\bar{b}}}-\bgo{\epsilon} \ge 1-\bgo{\epsilon}$ by Eq.~(\ref{eq:fourierapprox1}) and $\gamma=\tht{\epsilon}$), the outcome is $0^m$ and we obtain the state $\frac{h(A) \ket{\bar{b}}}{\nm{h(A) \ket{\bar{b}}}}$, which is $\epsilon$-close to $\ket{\bar{x}}$ in $l^2$ norm by Eq.~(\ref{eq:fourierapprox2}). To raise the success probability to $\omg{1}$, we use the standard ampltitude amplification \cite{BHMT00}, which requires 
\beba
\bgo{\dfrac{1}{\sqrt{p}}}=\bgo{\alpha}=\bgo{\kappa \sqrt{\log{\kappa/\epsilon}}}
\eeea
repetitions of the above procedure. This means that we need to implement each $U$ with precision 
\beba
\epsilon' =\bgo{\dfrac{\epsilon}{\alpha}}=\bgo{\dfrac{\epsilon}{\kappa \sqrt{\log{\kappa/\epsilon}}}}.
\eeea
The resulting algorithm, denoted by $\mcal{A}$, makes $$\bgo{d\kappa^2 \cdot \poly{\log{\dfrac{d \kappa}{\epsilon}}}}$$ uses of $\mcal{P}_A$ and $\mcal{P}_b$, and is gate-efficient. The $\kappa$-dependence can be decreased from quadratic to nearly linear by using  Ambainis' variable-time amplitude amplification \cite{Amb10}. However, we do not need this technique to prove Lemma \ref{lem:qlssolnom}, \ref{lem:qlssolentry} or \ref{lem:qlssolentrydiff}.

\proof{[Proof of Lemma \ref{lem:qlssolnom}] Suppose $||\vec{b}||=q$ is known. Then $\nm{\vec{x}}= q\nm{A^{-1}\ket{\bar{b}}}$. So in order to estimate $\nm{\vec{x}}$ up to multiplicative error $\bgo{\epsilon}$, we only need to get an $\bgo{\epsilon}$-multiplicative approximation of $\nm{A^{-1}\ket{\bar{b}}}$. To achieve this, we modify $\mcal{A}$ by replacing amplitude amplification with amplitude estimation \cite{BHMT00}. Specifically, we still choose $\gamma = \tht{\epsilon}$ and define the corresponding operators $V$, $U$ and $W$ as before. Then, if we measure the first $m$ qubits of $W \ket{0^m} \ket{b}$ in the standard basis, the probability of getting outcome $0^m$ is
\beba
p = \dfrac{\nm{h(A)\ket{\bar{b}}}^2}{\alpha^2} = \omg{\dfrac{1}{\alpha^2}},
\eeea
where $\alpha = \tht{\kappa \sqrt{\log{\kappa/\epsilon}}}$. We use 
amplitude estimation to obtain an $\bgo{\epsilon}$-multiplicative approximation $\hat{p}$ of $p$ (succeeding with probability at least $3/4$). Then $\sqrt{\hat{p}}$ is an $\bgo{\epsilon}$-multiplicative approximation of $\sqrt{p}$ (note that $1-\delta \le \sqrt{1-\delta} \le \sqrt{1+\delta} \le 1+\delta$ for all $\delta \in [0, 1]$), and hence $\alpha \sqrt{\hat{p}}$ is an $\bgo{\epsilon}$-multiplicative approximation of $\nm{h(A)\ket{\bar{b}}}$. Meanwhile, by Eq.~(\ref{eq:fourierapprox1}), $\gamma=\tht{\epsilon}$ and $\nm{A^{-1}\ket{\bar{b}}} \ge 1$, we know that $\nm{h(A)\ket{\bar{b}}}$ is an $\bgo{\epsilon}$-multiplicative approximation of $\nm{A^{-1}\ket{\bar{b}}}$. Combining these two facts, we get that $\alpha \sqrt{\hat{p}}$ is an $O(\epsilon)$-multiplicative approximation of $\nm{A^{-1}\ket{\bar{b}}}$. Therefore, $q\alpha \sqrt{\hat{p}}$ is an $\bgo{\epsilon}$-multiplicative approximation of $\nm{\vec{x}}$, as desired.

Let us analyze the complexity of this algorithm. Since we want to estimate $p=\omg{1/\alpha^2}$ up to multiplicative error $\bgo{\epsilon}$, amplitude estimation requires 
\beba
\bgo{\dfrac{1}{\epsilon \sqrt{p}}} = \bgo{\dfrac{\alpha}{\epsilon}}=\bgo{\dfrac{\kappa \sqrt{\log{\kappa/\epsilon}}}{\epsilon}}
\eeea
repetitions of $W$ and $\mcal{P}_b$. This means that we need to implement each $U$ with precision 
\beba
\bgo{\dfrac{\epsilon}{\alpha}}=\bgo{\dfrac{\epsilon}{\kappa \sqrt{\log{\kappa/\epsilon}}}}.
\eeea
This can be achieved by a gate-efficient procedure that makes $O(d \kappa \cdot \poly{\log{d \kappa/\epsilon}} )$ uses of $\mcal{P}_A$. So the resulting algorithm makes $$\bgo{\dfrac{d \kappa^2}{\epsilon} \cdot \poly{\log{\dfrac{d \kappa}{\epsilon}}}}$$ uses of $\mcal{P}_A$ and $\mcal{P}_b$, and is gate-efficient. 
}

\proof{[Proof of Lemma \ref{lem:qlssolentry}] Let $\ket{y}={A^{-1}\ket{\bar{b}}}$. Then $\abs{x_i} =  q \abs{\braket{i}{y}}$. So in order to estimate $\abs{x_i}$ up to additive error $\bgo{\epsilon}$, 
we only need to get an $\bgo{\epsilon/q}$-additive approximation of $\abs{\braket{i}{y}}$. To achieve this, we choose $\gamma=\tht{\epsilon'}$ where $\epsilon' \defeq \epsilon/q$, so that
\beba
\nm{h(A) \ket{\bar{b}}  - A^{-1}\ket{\bar{b}}} =\bgo{\epsilon'}.
\eeea
Let $\ket{y'}=h(A) \ket{\bar{b}}$. Then $\nm{\ket{y'}-\ket{y}} =\bgo{\epsilon'}$. Next, we define the operators $V$, $U$ and $W$ corresponding to this $\gamma$. We also define a unitary operator $R$ such that $R\ket{0}\ket{i} = \ket{1}\ket{i}$ and $R\ket{0}\ket{i'}=\ket{0}\ket{i'}$ for all $i' \neq i$. Then we have
\beba
RW \ket{0^m}\ket{0} \ket{\bar{b}}
=\dfrac{\braket{i}{y'}}{\alpha} \ket{0^m} \ket{1} \ket{i} 
+\dsum_{i' \neq i} \dfrac{\braket{i'}{y'}}{\alpha} \ket{0^m} \ket{0} \ket{i'}
+\ket{\Xi^{\perp}},
\eeea
where $\alpha=\tht{\kappa \sqrt{\log{\kappa/\epsilon'}}}$, and $\ket{\Xi^{\perp}}$ is an unnormalized state satisfying 
$(\ketbra{0^m}{0^m}\otimes I) \ket{\Xi^{\perp}}=0$. Then, if we measure the first $m+1$ qubits of this state in the standard basis, the probability of getting  outcome $0^m 1$ is 
\beba
p' \defeq \dfrac{\abs{\braket{i}{y'}}^2}{\alpha^2}.
\eeea
We use amplitude estimation to obtain an $\bgo{\epsilon''}$-additive approximation $\hat{p}'$ of $p'$  (succeeding with probability at least $3/4$), where $\epsilon'' \defeq (\epsilon')^2 /\alpha^2$.
Then $\sqrt{\hat{p}'}$ is an $\bgo{\sqrt{\epsilon''}}$-additive approximation of
$\sqrt{p'}$ (note that $\sqrt{z}-\sqrt{\delta} \le \sqrt{z-\delta} \le \sqrt{z+\delta} \le \sqrt{z}+\sqrt{\delta}$ for all $z \ge \delta \ge 0$), and hence
\beba
\abs{\abs{\braket{i}{y'}} - \alpha \sqrt{\hat{p}'}}
=\abs{\alpha \sqrt{p'} - \alpha \sqrt{\hat{p}'}}
=\bgo{ \alpha \sqrt{\epsilon''}}  
=\bgo{\epsilon'}.
\label{eq:apiy}
\eeea
Meanwhile, note that 
\beba
\abs{\braket{i}{y'}-\braket{i}{y}} \le \nm{\ket{y'}-\ket{y}} =\bgo{\epsilon'}.
\label{eq:iyiy}
\eeea
Combining Eqs.~(\ref{eq:apiy}) and (\ref{eq:iyiy}), we get
\beba
\abs{\alpha \sqrt{\hat{p}'} - \abs{\braket{i}{y}}} = \bgo{\epsilon'}.
\eeea
Then, since $\epsilon=q \epsilon'$, we know that $q\alpha \sqrt{\hat{p}'}$ is an $\bgo{\epsilon}$-additive approximation of $\abs{x_i}=q \abs{\braket{i}{y}}$, as desired.

Let us analyze the complexity of this algorithm. Since we want to estimate $p'$ up to additive error $\bgo{\epsilon''}$ where $\epsilon'' =\tht{\epsilon^2/(q^2 \kappa^2 \log{q \kappa/\epsilon})}$, amplitude estimation requires 
\beba
\bgo{\dfrac{1}{\epsilon''}}=\bgo{\dfrac{q^2 \kappa^2 \log{q \kappa/\epsilon}}{\epsilon^2}}
\eeea
repetitions of $R$, $W$ and $\mcal{P}_b$. This means that we need to implement each $U$ with precision 
\beba
\bgo{\epsilon''}=\bgo{\dfrac{\epsilon^2}{q^2 \kappa^2 \log{q \kappa/\epsilon}}}.
\eeea
This can be achieved by a gate-efficient procedure that makes $\bgo{d \kappa \cdot \poly{\log{d \kappa q/\epsilon}} }$ uses of $\mcal{P}_A$. So the resulting algorithm makes $$\bgo{\dfrac{d q^2 \kappa^3}{\epsilon^2} \cdot \poly{\log{\dfrac{dq \kappa }{\epsilon}}}}$$ uses of $\mcal{P}_A$ and $\mcal{P}_b$, and is gate-efficient. 
}

\proof{[Proof of Lemma \ref{lem:qlssolentrydiff}] The proof of this lemma is similar to that of Lemma \ref{lem:qlssolentry}. The main difference is that here we replace $R$ with a unitary operation $Q$ which satisfies $Q \ket{0} \ket{-_{i,j}} = \ket{1} \ket{-_{i,j}} $, $Q \ket{0} \ket{+_{i,j}} = \ket{0} \ket{+_{i,j}} $, and $Q \ket{0}  \ket{l} = \ket{0} \ket{l}$ for all $l \neq i, j$, where $\ket{\pm_{i, j}} \defeq \lb \ket{i} \pm \ket{j}\rb/\sqrt{2}$. Then we have
\bea
QW \ket{0^m}\ket{0} \ket{\bar{b}}
&=&\dfrac{\braket{-_{i, j}}{y'}}{\alpha} \ket{0^m} \ket{1} \ket{-_{i,j}}
+\dfrac{\braket{+_{i, j}}{y'}}{\alpha} \ket{0^m} \ket{0} \ket{+_{i,j}} \\
&&+\dsum_{l \neq i,j} \dfrac{\braket{l}{y'}}{\alpha} \ket{0^m} \ket{0} \ket{l}
+\ket{\Xi^{\perp}},
\eea
where $W$ is defined as in the proof of Lemma \ref{lem:qlssolentry}, $\ket{y'}=h(A) \ket{\bar{b}}$, and $\ket{\Xi^{\perp}}$ is an unnormalized state satisfying 
$(\ketbra{0^m}{0^m}\otimes I) \ket{\Xi^{\perp}}=0$. We still pick $\gamma=\tht{\epsilon/q}$ such that $\nm{\ket{y}-\ket{y'}} =\bgo{\epsilon/q}$, where
$\ket{y}=A^{-1}\ket{\bar{b}}$. If we measure the first $m+1$ qubits of $QW \ket{0^m}\ket{0} \ket{\bar{b}}$, then the probability of getting outcome $0^m1$ is 
\beba
p' \defeq \dfrac{\abs{\braket{-_{i,j}}{y'}}^2}{\alpha^2}.
\eeea
We use amplitude estimation to obtain an $\bgo{\epsilon^2/(q^2 \alpha^2)}$-additive approximation $\hat{p}'$ of $p'$ (succeeding with probability at least $3/4$). Then $\sqrt{\hat{p}'}$ is an $\bgo{\epsilon/(q \alpha)}$-additive approximation of $\sqrt{p'}$, and hence $ \alpha \sqrt{\hat{p}'}$ is an $\bgo{\epsilon/q}$-additive approximation of $\alpha \sqrt{p'}=\abs{\braket{-_{i,j}}{y'}}$. Meanwhile,  
\beba
\abs{\braket{-_{i.j}}{y'} - \braket{-_{i,j}}{y}}
\le \nm{ \ket{y'} - \ket{y}}
=\bgo{\dfrac{\epsilon}{q}},
\eeea 
Combining these two facts, we know that $ q \alpha \sqrt{\hat{p}'}$ is an $\bgo{\epsilon}$-additive approximation of 
\beba
q\abs{\braket{-_{i,j}}{y}}=\dfrac{\abs{x_i-x_j}}{\sqrt{2}}.
\eeea
All the parameters are on the same order as in the proof of Lemma \ref{lem:qlssolentry}. So this algorithm also makes $$\bgo{\dfrac{d q^2 \kappa^3}{\epsilon^2} \cdot \poly{\log{\dfrac{d q\kappa}{\epsilon}}}}$$ uses of $\mcal{P}_A$ and $\mcal{P}_b$, and is gate-efficient. 
}

It is worth noting that the algorithms in Theorem \ref{thm:qlssol} and Lemmas \ref{lem:qlssolnom}, \ref{lem:qlssolentry}, \ref{lem:qlssolentrydiff} still work when $A$ is not invertible but $\vec{b} \in \range{A}$. In this case, we only need to replace $A^{-1}$ with $A^+$, and replace the condition number $\kappa$ of $A$ with the finite condition number $\kappa_f$ of $A$. This property will be useful in the next subsection.

\subsection{Using QLSAs to analyze electrical networks}
\label{sec:aenbyqlsa}
Now we show how to use the QLSAs in Lemmas \ref{lem:qlssolnom}, \ref{lem:qlssolentry} and \ref{lem:qlssolentrydiff} to analyze electrical networks. Let $G=(V, E, w)$ be an electrical network driven by an external current $\mbf{i}_{ext}$, where $\card{V}=N$, $\uwdg{G}=d$, $\lambda_2(\nlap) \ge \lambda >0$, $1 \le w_e \le c$ for all $e \in E$, $\mbf{i}_{ext} \perp \mbf{1}$ and $\nm{\viext}=1$. Let $\mbf{v} \in \bbR^V$ be the induced potentials at the vertices, and let $\mbf{i} \in \bbR^E$ be the induced currents on the edges. To solve the ENA-V problem, we consider the Laplacian system
\beba
\lap \mbf{v} = \mbf{i}_{ext}.
\eeea

\begin{theorem}
The ENA-V problem can be solved by a gate-efficient quantum algorithm that makes 
$$\bgo{\dfrac{c d^3 }{\lambda^3 \epsilon^2} \cdot \poly{\log{\dfrac{cd}{\lambda \epsilon}}}}$$ 
uses of $\mcal{P}_v$, $\mcal{P}_e$ and $\mcal{P}_i$. 
\label{thm:enav1}
\end{theorem} 
\proof{ For any $v \in V$, we have $1 \le \wdg{v}=\sum_{e \in E(v)} w_e \le cd$. So by Eq.~(\ref{eq:eigvallap}) and Lemma \ref{lem:lapnlapspectralgap}, 
we have
\beba
0=\lambda_1(\lap) < \lambda \le \lambda_2(\lap) \le \dots \le \lambda_N(\lap) \le 2cd.
\eeea
Let $A \defeq \frac{1}{2cd}\lap$ and $\vec{b} \defeq \frac{1}{2cd}\viext$. Then we get
\beba
\mbf{v} = \lap^+ \mbf{i}_{ext} = A^+ \vec{b}.
\eeea
Note that all the nonzero eigenvalues of $A$ are in the range $[\lambda/(2cd), 1]$, which means that $A$ has finite condition number $\kappa_f \le 2cd/\lambda$. In addition, $A$ is $(d+1)$-sparse, and given any $(i, j) \in \intset{N} \times \intset{d+1}$, we can find the location and value of the $j$-th nonzero entry in the $i$-th row of $A$ by making $\bgo{d}$ uses of $\mcal{P}_v$ and $\mcal{P}_e$. Meanwhile, we have $\vec{b} \in \range{A}$, $g \defeq ||\vec{b}||= 1/(2cd)$, and we can prepare a state proportional to $\vec{b}$ by calling $\mcal{P}_i$ once. Using these facts and Lemma \ref{lem:qlssolentrydiff}, we know that there exists a gate-efficient quantum algorithm that estimates $\abs{\mbf{v}(s)-\mbf{v}(t)}$ up to additive error $\epsilon$, for any given $s, t \in V$, by making
\beba
\bgo{d \cdot \dfrac{d g^2 \kappa_f ^3}{\epsilon^2} \cdot \poly{\log{\dfrac{d g \kappa_f }{ \epsilon}}}}
=\bgo{\dfrac{c d^3 }{\lambda^3 \epsilon^2} \cdot \poly{\log{\dfrac{cd}{\lambda \epsilon}}}}
\eeea
uses of $\mcal{P}_v$, $\mcal{P}_e$ and $\mcal{P}_i$, as claimed.
}

To solve the ENA-C, ENA-P and ENA-ER problems, we consider another linear system. This system has the advantage that the finite condition number of its coefficient matrix is the square root of that of Laplacian system. As a result, it can be solved more efficiently. Recall that $B_G \mbf{i}=\mbf{i}_{ext}$ and $C_G=B_G W_G^{1/2}$. So we have
\beba
\bpm
0 & C_G \\
C_G^T & 0 
\epm
\bpm
0 \\
W_G^{-1/2}\mbf{i}
\epm
=\bpm
\mbf{i}_{ext}\\
0
\epm
\eeea
Furthermore, we claim:

\begin{lemma}
\beba
\bpm
0 & C_G \\
C_G^T & 0 
\epm^+ \bpm
\mbf{i}_{ext}\\
0
\epm =  \bpm
0 \\
W_G^{-1/2}\mbf{i}
\epm.
\eeea
\label{lem:linearsystem2}
\end{lemma}
\proof{ Suppose $C_G$ has the singular value decomposition $C_G=\sum_j s_j \ketbra{u_j}{v_j}$, where $s_j>0$, $\ket{u_j}$ and $\ket{v_j}$ are real vectors, for all $j$. Then $\lap=C_GC_G^T$ has the spectral decomposition 
$\lap=\sum_j s_j^2 \ketbra{u_j}{u_j}$. Moreover, since $\mbf{i}_{ext} \in  \range{\lap}$, we have $\mbf{i}_{ext}=\sum_j \alpha_j \ket{u_j}$ for some numbers $\alpha_j$'s. It follows that 
\beba
W_G^{-1/2}\ket{\mbf{i}}
=  W_G^{1/2}B_G^T\lap^+ \ket{\mbf{i}_{ext} } 
=  C_G^T \lap^+ \ket{\mbf{i}_{ext}} 
= \dsum_j s_j^{-1} \alpha_j \ket{v_j}.
\eeea
Meanwhile, we have
\bea
\bpm
0 & C_G \\
C_G^T & 0 
\epm
&=&\ketbra{1}{0} \otimes C_G^T
+\ketbra{0}{1} \otimes C_G \\
&=&\dsum_j s_j (\ketbra{1}{0} \otimes \ketbra{v_j}{u_j}
+\ketbra{0}{1} \otimes \ketbra{u_j}{v_j}) \\
&=& \dsum_j s_j \ketbra{+_j}{+_j}
-\dsum_j s_j \ketbra{-_j}{-_j},
\eea 
where 
\beba
\ket{\pm_j} \defeq \dfrac{\ket{1}\ket{v_j}\pm \ket{0}\ket{u_j}}{\sqrt{2}}.
\eeea
We also have
\beba
\bpm
\mbf{i}_{ext}\\
0
\epm
= \ket{0} \ket{\mbf{i}_{ext}} 
= \dsum_j \alpha_j \ket{0} \ket{u_j} 
= \dsum_j \dfrac{\alpha_j}{\sqrt{2}} \lb \ket{+_j}-\ket{-_j} \rb.
\eeea
These facts imply that
\bea
\bpm
0 & C_G \\
C_G^T & 0 
\epm^+ \bpm
\mbf{i}_{ext}\\
0
\epm
&=& \dsum_j  \dfrac{\alpha_j s_j^{-1}}{\sqrt{2}} \lb \ket{+_j}+\ket{-_j} \rb\\
&=& \dsum_j \alpha_j s_j^{-1} \ket{1} \ket{v_j} \\
&=& \bpm
0 \\
W_G^{-1/2}\mbf{i}
\epm.
\eea
}

\begin{theorem}
The ENA-C problem can be solved by a gate-efficient quantum algorithm that makes 
$$\bgo{\dfrac{c^{1.5} d^{1.5}}{\lambda^{1.5}\epsilon^2} \cdot \poly{\log{\dfrac{cd}{\lambda\epsilon}}}}$$
 uses of $\mcal{P}_v$, $\mcal{P}_e$ and $\mcal{P}_i$. 
\label{thm:enac1}
\end{theorem} 
\proof{ Let $A \defeq \frac{1}{\sqrt{2cd}}(\ketbra{1}{0} \otimes C_G^T
+\ketbra{0}{1} \otimes C_G)$ \footnote{Here we consider $A$ as a $2\card{E}\times 2\card{E}$ matrix with $\card{E}+\card{V}$ nonzero rows and 
$\card{E}+\card{V}$ nonzero columns.}
and $\ket{b} \defeq \frac{1}{\sqrt{2cd}}\ket{0} \ket{\mbf{i}_{ext}}$. 
Then Lemma \ref{lem:linearsystem2} implies
\beba
\ket{x} \defeq \ket{1} (W_G^{-1/2}\ket{\mbf{i}}) = A^+ \ket{b}.
\eeea
Thus, for any given $e \in E$, we have 
\beba
\abs{\mbf{i}(e)}=\sqrt{w_e} \abs{\braket{1,e}{x}},
\eeea
where $\ket{1,e} \defeq \ket{1}\ket{e}$. So in order to estimate $\abs{\mbf{i}(e)}$ up to additive error $\epsilon$, we only need to get an $\epsilon'$-additive approximation of $\abs{\braket{1,e}{x}}$, where $\epsilon' \defeq \epsilon/\sqrt{c}$, since $1 \le w_e \le c$.

By the proof of Theorem \ref{thm:enav1}, we know that all the nonzero eigenvalues of $\lap$ are in the range $[\lambda, 2cd]$. Then by the proof of Lemma \ref{lem:linearsystem2}, we know that all the nonzero singular values of $C_G$ are in the range $[\sqrt{\lambda}, \sqrt{2cd}]$, and all the nonzero eigenvalues of $A$ are in the range $[-1, -\sqrt{\lambda/(2cd)}] \cup [\sqrt{\lambda/(2cd)}, 1]$. This implies that $A$ has finite condition number $\kappa_f \le \sqrt{2cd/\lambda}$. Moreover, $A$ is $d$-sparse, and for any given $(i, j) \in \intset{2\card{E}} \times \intset{d}$, we can find the location and value of the $j$-th nonzero entry in the $i$-th row of $A$ by making $\bgo{1}$ uses of $\mcal{P}_v$ and $\mcal{P}_e$. Furthermore, by the proof of Lemma \ref{lem:linearsystem2}, we know that $\ket{b} \in \range{A}$. We also have $g \defeq \nm{\ket{b}}=1/\sqrt{2cd}$, and we can prepare a state proportional to $\ket{b}$ by calling $\mcal{P}_i$ once. Using these facts and Lemma \ref{lem:qlssolentry}, we know that there exists a gate-efficient quantum algorithm that estimates $\abs{\braket{1,e}{x}}$ up to additive error $\epsilon'$ by making
\beba
\bgo{\dfrac{d g^2 \kappa_f^3}{(\epsilon')^2}\cdot \poly{\log{\dfrac{dg\kappa_f}{\epsilon'}}}}
=
\bgo{\dfrac{c^{1.5} d^{1.5}}{\lambda^{1.5}\epsilon^2} \cdot \poly{\log{\dfrac{cd}{\lambda\epsilon}}}}
\eeea
uses of $\mcal{P}_v$, $\mcal{P}_e$ and $\mcal{P}_i$. This concludes the proof.
}

Theorem \ref{thm:enac1} also provides a method for estimating the voltage between two \emph{adjacent} vertices $s$ and $t$:

\begin{corollary}
Under the promise that $s$ and $t$ are adjacent vertices, the ENA-V problem can be solved by a gate-efficient algorithm that makes 
$$\bgo{\dfrac{c^{1.5} d^{1.5} }{\lambda^{1.5}\epsilon^2} \cdot \poly{\log{\dfrac{cd}{\lambda\epsilon}}}}$$ 
uses of $\mcal{P}_v$, $\mcal{P}_e$ and $\mcal{P}_i$.
\label{cor:enav2}
\end{corollary}
\proof{ Let $e=(s, t) \in E$. By Ohm's law, we have $\abs{\mbf{v}(s)-\mbf{v}(t)} = \abs{\mbf{i}(e)}/w_e$, where $1 \le w_e \le c$. So in order to estimate $\abs{\mbf{v}(s)-\mbf{v}(t)}$ up to additive error $\epsilon$, we only need to get an $\epsilon$-additive approximation of $\abs{\mbf{i}(e)}$. By Theorem \ref{thm:enac1}, this can be achieved by a gate-efficient quantum algorithm that makes $O((c^{1.5} d^{1.5} /(\lambda^{1.5}\epsilon^2)) \cdot \poly{\log{cd/(\lambda\epsilon)}})$ uses of $\mcal{P}_v$, $\mcal{P}_e$ and $\mcal{P}_i$. 
}

One can compare the algorithm in Corollary \ref{cor:enav2} with the one in Theorem \ref{thm:enav1} for solving ENA-V in the general case. The former has better dependence on $d$ and $1/\lambda$, but slightly worse dependence on $c$, than the latter. So they are incomparable.

Finally, we solve the ENA-P and ENA-ER problems by utilizing the algorithm in Lemma \ref{lem:qlssolnom}: 

\begin{theorem}
The ENA-P problem can be solved by a gate-efficient quantum algorithm that makes 
$$\bgo{\dfrac{c d^2}{\lambda \epsilon} \cdot \poly{\log{\dfrac{c d }{\lambda \epsilon}}}}$$ 
uses of $\mcal{P}_v$, $\mcal{P}_e$ and $\mcal{P}_i$. 
\label{thm:enap1}
\end{theorem} 
\proof{ Let us use the same notation as in the proof of Theorem \ref{thm:enac1}. Then we have 
\beba
\nm{\ket{x}}^2 = \nm{W_G^{-1/2} \ket{\mbf{i}}} = \bra{\vi} W_G^{-1} \ket{\vi}=\mcal{E}(\mbf{i}).
\eeea
Thus, in order to estimate $\mcal{E}(\mbf{i})$ up to multiplicative error $\bgo{\epsilon}$, we only need to get an $\bgo{\epsilon}$-multiplicative approximation of $\nm{\ket{x}}$ (note that $1-2\delta \le (1-\delta)^2 \le (1+\delta)^2 \le 1+3\delta$ for all $\delta \in [0, 1]$). By Lemma \ref{lem:qlssolnom}, this can be accomplished by a gate-efficient quantum algorithm that makes
\beba
\bgo{\dfrac{d \kappa_f^2}{\epsilon} \cdot \poly{\log{\dfrac{d \kappa_f}{\epsilon}}}}
=\bgo{\dfrac{c d^2}{\lambda \epsilon} \cdot \poly{\log{\dfrac{c d }{\lambda \epsilon}}}}
\eeea
uses of $\mcal{P}_v$, $\mcal{P}_e$ and $\mcal{P}_i$. This concludes the proof.
}

\begin{corollary}
The ENA-ER problem can be solved by a gate-efficient quantum algorithm that makes $$\bgo{\dfrac{c d^2}{\lambda \epsilon} \cdot \poly{\log{\dfrac{c d }{\lambda \epsilon}}}}$$ uses of $\mcal{P}_v$ and $\mcal{P}_e$.
\label{cor:enaer1}
\end{corollary}
\proof{ Recall that $\efr(s, t) = \mcal{E}(\mbf{i})$, where $\mbf{i}=W_GB_G^T\lap^+ \chi_{s,t}$ is the electrical flow induced by the external current $\chi_{s,t}=\ket{s}-\ket{t}$. Clearly, we can prepare the state $(\ket{s}-\ket{t})/\sqrt{2}$ in time $\poly{\log{N}}$. Then, by Theorem \ref{thm:enap1}, there is a gate-efficient algorithm that makes $O((c d^2/(\lambda \epsilon)) \cdot \poly{\log{c d /(\lambda \epsilon)}})$ uses of $\mcal{P}_v$ and $\mcal{P}_e$, and outputs an $\bgo{\epsilon}$-multiplicative approximation of $\mcal{E}(\tilde{\vi})=\mcal{E}(\vi)/2$, where $\tilde{\vi}=\vi/\sqrt{2}$ is the electrical flow induced by the external current $\chi_{s,t}/\sqrt{2}$. Then we multiply this result by a factor of $2$, and obtain an $\bgo{\epsilon}$-multiplicative approximation of $\mcal{E}(\vi)=\efr(s, t)$.
}

\section{Analyzing Electrical Networks by Using Quantum Walks}
\label{sec:algorithmsquantumwalk}

In this section, we present a set of quantum algorithms for analyzing electrical networks based on using quantum walks. In Section \ref{sec:modifygraphdefinematrix}, we define several graph-related matrices, and show that they have some nice properties. In Section \ref{sec:definequantumwalk}, we quantize one of the matrices to obtain a quantum walk. Then we analyze the spectral properties of this quantum walk, and give its efficient implementations. In Section \ref{sec:aenbyqw}, we describe how to utilize this quantum walk to solve the ENA-P and ENA-ER problems.

\subsection{Modifying graphs and defining matrices}
\label{sec:modifygraphdefinematrix}

Suppose $G=(V, E, w)$ is an electrical network driven by an external current $\mbf{i}_{ext}$, where $\card{V}=N$, $\uwdg{G}=d$, $\lambda_2(\nlap) \ge \lambda >0$, $1 \le w_e \le c$ for all $e \in E$, $\mbf{i}_{ext} \perp \mbf{1}$ and $\nm{\viext}=1$. For any $e \in E$, let 
\bea
\ket{\vphi_e} \defeq \sqrt{w_e} ( \ket{e^-}-\ket{e^+}).
\eea
Then we have
\beba
C_G=B_G W_G^{1/2} = \dsum_{e \in E} \ket{\vphi_e}\bra{e}.
\eeea

Now we modify the graph $G$ as follows. We add a special hyperedge $e_0$ among all the vertices in $V$, and set its weight to be $\lambda$. Let $G'=(V,E',w)$ be the modified hypergraph, where $E' \defeq E \cup \seta{e_0}$, $w_e$ is the same as before for all $e \in E$, and $w_{e_0}=\lambda$. Let 
\bea
\ket{\vphi_{e_0}} \defeq  -\sqrt{2\lambda}\ket{\viext} = -\dsum_{v \in V} \sqrt{2\lambda} \viext(v) \ket{v}.
\eea
Then we define 
\beba
C_{G'} \defeq \dsum_{e\in E'} \ketbra{\vphi_e}{e} = C_G + \ketbra{\vphi_{e_0}}{e_0}.
\eeea
Moreover, for any $v \in V$, let $E'(v) \defeq E(v) \cup \seta{e_0}$, and let $\wdgp{v} \defeq \wdg{v}+\lambda$. Then we define
\beba
D_{G'} \defeq \dsum_{v \in V} \wdgp{v} \ketbra{v}{v}.
\eeea
We also define
\beba
\lapp \defeq C_{G'} C_{G'}^T,
\eeea
and
\beba
\nlapp \defeq D_{G'}^{-1/2} \lapp D_{G'}^{-1/2}.
\eeea
One can easily see that
\bea
\ker{\lapp}&=&\spn{\ket{\mbf{1}}}, \\
\ker{\nlapp}&=&\spn{D_{G'}^{1/2}\ket{\mbf{1}}},
\eea
and
\bea
\range{\lapp} &=& \seta{\ket{\psi}:~\ket{\psi} \perp \ket{\mbf{1}}}, \\
\range{\nlapp} &=& \seta{D_{G'}^{-1/2}\ket{\psi}:~\ket{\psi} \perp \ket{\mbf{1}}},
\eea
where $\ket{\mbf{1}}=\sum_{v} \ket{v}$.

The following lemma establishes a relationship between $\lambda_2(\nlap)$ and $\lambda_2(\nlapp)$:

\begin{lemma}
If $\lambda_2(\nlap) \ge \lambda > 0$, then $\lambda_2(\nlapp)\ge \lambda/3>0$.
\label{lem:nlapgnlapgpspectralgap}
\end{lemma}
\proof{ It is sufficient to show that, for any $\ket{\vphi} \in \range{\nlapp}$, $\braket{\vphi}{\vphi} \neq 0$, we have
\beba
\bra{\vphi} \nlapp \ket{\vphi} \ge \dfrac{\lambda}{3} \cdot \braket{\vphi}{\vphi}.
\eeea

Suppose $\ket{\vphi}= D_{G'}^{-1/2} \ket{\psi}$ for some $\ket{\psi} \perp \ket{\mbf{1}}$, $\braket{\psi}{\psi} \neq 0$. Then, since $D_G^{-1/2}\ket{\psi} \in \range{\nlap}$ and $\lambda_2(\nlap) \ge \lambda$, we get
\beba
\nlap 
\matge 
\lambda  \dfrac{D_G^{-1/2}\ketbra{\psi}{\psi}D_G^{-1/2}}{\bra{\psi} D_G^{-1}\ket{\psi}}.
\eeea
This implies that
\beba
D_G^{1/2}\nlap D_G^{1/2}
\matge
\lambda  \dfrac{\ketbra{\psi}{\psi}}{\bra{\psi} D_G^{-1}\ket{\psi}}.
\label{eq:dglgdg}
\eeea
Meanwhile, note that
\beba
\lapp = C_{G'}C_{G'}^T
=\dsum_{e \in E'} \ketbra{\vphi_e}{\vphi_e}
\matge 
\dsum_{e \in E} \ketbra{\vphi_e}{\vphi_e}
= C_G C_G^T = \lap.
\eeea
This implies that
\beba
\nlapp=D_{G'}^{-1/2} \lapp D_{G'}^{-1/2}
 \matge D_{G'}^{-1/2} \lap D_{G'}^{-1/2}
 =D_{G'}^{-1/2}D_{G}^{1/2} \nlap D_{G}^{1/2} D_{G'}^{-1/2}.
 \label{eq:lgplg}
 \eeea
Combining Eqs.~(\ref{eq:dglgdg}) and (\ref{eq:lgplg}), we get
\bea
\bra{\vphi} \nlapp \ket{\vphi}
&\ge&
\bra{\vphi}D_{G'}^{-1/2}D_{G}^{1/2} \nlap D_{G}^{1/2} D_{G'}^{-1/2}
\ket{\vphi} \\
&=&
\bra{\psi}D_{G'}^{-1}D_{G}^{1/2} \nlap D_{G}^{1/2} D_{G'}^{-1}
\ket{\psi} \\
&\ge&
\lambda \dfrac{\bra{\psi}D_{G'}^{-1}\ketbra{\psi}{\psi} D_{G'}^{-1}
\ket{\psi}}{\bra{\psi} D_G^{-1}\ket{\psi}} \\
&\ge &
\lambda \braket{\vphi}{\vphi}
\dfrac{\bra{\psi} D_{G'}^{-1}
\ket{\psi}}{\bra{\psi} D_G^{-1}\ket{\psi}}.
\label{eq:philgphi}
\eea
Suppose $\ket{\psi} = \sum_{v \in V} \alpha_v \ket{v}$ for some numbers $\alpha_v$'s. Then
\beba
\dfrac{\bra{\psi} D_{G'}^{-1}
\ket{\psi}}{\bra{\psi} D_G^{-1}\ket{\psi}}
=\dfrac{\dsum_{v \in V} \wdgp{v}^{-1} \abs{\alpha_v}^2 }{\dsum_{v \in V}  \wdg{v}^{-1} \abs{\alpha_v}^2}.
\eeea
For any $v \in V$, we have $1\le \wdg{v} \le cd$. Then, since $\lambda \le \lambda_2(\nlap) \le 2$, we have
\beba
\dfrac{\wdg{v}}{\wdgp{v}}=\dfrac{\wdg{v}}{\wdg{v}+\lambda}
\ge \dfrac{1}{1+\lambda}
\ge \dfrac{1}{3}.
\eeea
It follows that
\beba
\dfrac{\bra{\psi} D_{G'}^{-1}
\ket{\psi}}{\bra{\psi} D_G^{-1}\ket{\psi}}
\ge \min\limits_{v \in V} \dfrac{\wdgp{v}^{-1}}{\wdg{v}^{-1}}
\ge \dfrac{1}{3}.
\eeea
Plugging this into Eq.~(\ref{eq:philgphi}) yields
\beba
\bra{\vphi} \nlapp \ket{\vphi} \ge \dfrac{\lambda}{3} \cdot \braket{\vphi}{\vphi},
\eeea
as desired.
}

Now let us consider $\ker{C_{G'}}$. We claim:

\begin{lemma}
\beba
\ker{C_{G'}}=\seta{g(\mbf{f}):~\mbf{f} \mrm{~is~a~flow~consistent~with~} \alpha \cdot \mbf{i}_{ext} \mrm{~for~some~number~} \alpha}, 
\eeea
where
\beba
g(\mbf{f}) \defeq \dfrac{1}{\sqrt{2\lambda}}h(\vf)\ket{e_0}+\dsum_{e \in E} \dfrac{\mbf{f}(e)}{\sqrt{w_e}}\ket{e},
\label{eq:form}
\eeea
in which $h(\vf)=\alpha$ if $\mbf{f}$ is consistent with $\alpha \cdot \viext$.
\label{lem:kercg}
\end{lemma}
\proof{ Let $\ket{\psi} = \beta \ket{e_0} + \sum_{e \in E} \beta_e \ket{e}$ be arbitrary. Then 
\beba
C_{G'} \ket{\psi} = \beta \ket{\vphi_{e_0}} +
\dsum_{e \in E}  \beta_e \ket{\vphi_e}=0
\eeea
if and only if
\beba
\dsum_{e \in E^-(v)} \sqrt{w_e} \beta_e
-\dsum_{e \in E^+(v)} \sqrt{w_e} \beta_e
= \sqrt{2\lambda} \beta \cdot \viext(v), & \forall v \in V.
\eeea
This is equivalent to the condition that the flow $\vf$ defined as $\vf(e)= \sqrt{w_e} \beta_e$ for all $e \in E$ is consistent with $\sqrt{2\lambda}  \beta \cdot \viext$. 
} 

Now let $\Pi$ be the projection onto $\ker{C_{G'}}$. We claim:

\begin{lemma}
\beba
\Pi \ket{e_0} \propto \ket{\Phi} \defeq g(\vi) = 
\dfrac{1}{\sqrt{2\lambda}} \ket{e_0}+\dsum_{e \in E} \dfrac{\mbf{i}(e)}{\sqrt{w_{e}}}\ket{e},
\eeea
where $\vi=W_GB_G^T\lap^+\viext$ is the electrical flow induced by $\viext$.
\label{lem:projkercg}
\end{lemma}

We will give two proofs of Lemma \ref{lem:projkercg}. The first one is algebraic and more rigorous, and the second one is geometric and more intuitive.

\proof{[Proof 1 of Lemma \ref{lem:projkercg}] It is sufficient to show that for any $\ket{\Psi} \in \ker{C_{G'}}$, if $\ket{\Psi} \bot \ket{\Phi}$, then $\ket{\Psi} \bot \ket{e_0}$. We prove its contrapositive by contradiction.

Suppose $\ket{\Psi} \in \ker{C_{G'}}$ satisfies $\ket{\Psi} \not\perp \ket{e_0}$. By Lemma \ref{lem:kercg}, after appropriate rescaling of $\ket{\Psi}$, we can write it as
\beq
\ket{\Psi}=g(\vf)=\dfrac{1}{\sqrt{2\lambda}}\ket{e_0}+\sum_{e \in E} {\dfrac{\vf(e)}{\sqrt{w_{e}}}}\ket{e}
\eeq
for some flow $\vf$ consistent with $\viext$. We claim that, if
\beq
\ket{\Psi} \perp \ket{\Phi}
=g(\vi) =\dfrac{1}{\sqrt{2\lambda}} \ket{e_0}+\dsum_{e \in E} \dfrac{\mbf{i}(e)}{\sqrt{w_{e}}}\ket{e},
\eeq 
which means that
\beq
- \dfrac{1}{2\lambda}=\dsum_{e \in E} \dfrac{\vf(e)\vi(e)}{w_{e}}<0,
\eeq
then there exists a flow $\vf' \neq \vi$ consistent with $\viext$ such that $\mcal{E}(\vf')<\mcal{E}(\vi)$. But this is contradictory to Lemma \ref{lem:minpower} which states that $\vi$ has the minimum power among such  flows! Consequently, we must have $\ket{\Psi} \not\perp \ket{\Phi}$.

Now we prove this claim. Let $\vf' = \beta \vf + (1-\beta) \vi$, where $\beta \in (0,1)$ is to be chosen later. Obviously, $\vf'$ is a flow consistent with $\viext$ for any choice of $\beta$. Let us consider the power of $\vf'$.
\bea
\mcal{E}(\vf') &=& \dsum_{e \in E}\dfrac{(\beta \vf(e)+(1-\beta) \vi(e))^2}{w_{e}} \\
&=& \beta^2 \dsum_{e \in E} \dfrac{\vf(e)^2}{w_{e}}
+(1-\beta)^2 \dsum_{e \in E}\dfrac{\vi(e)^2}{w_{e}}
+2 \beta (1-\beta) \dsum_{e \in E}\dfrac{\vf(e) \vi(e)}{w_{e}} \\
&<& \beta^2 \mcal{E}(\vf)
+(1-\beta)^2 \mcal{E}(\vi).
\eea
Now let $\gamma = \dfrac{\mcal{E}(\vf)}{\mcal{E}(\vi)}$ and $\beta = \dfrac{1}{1+\gamma}$. Then we get
\bea
\mcal{E}(\vf') < (\beta^2 \gamma+(1-\beta)^2) \mcal{E}(\vi) 
= \dfrac{\gamma}{1+\gamma} \mcal{E}(\vi)
< \mcal{E}(\vi),
\eea
as claimed.
}

\proof{[Proof 2 of Lemma \ref{lem:projkercg}] Consider the geometric picture shown in Fig.\ref{fig:fig1}. 

\begin{figure}[H]
\begin{center}
\input{projection_fig.tex}
\end{center}
\fcaption{Geometric proof of Lemma \ref{lem:projkercg}. Here $L_1$ and $L_2$ are two hyperplanes in the space $\mcal{H}=\spn{\ket{e}: e \in E'}$. They are defined as $L_1= \ker{C_{G'}}$ and $L_2 = \seta{ \ket{\Psi} \in \mcal{H}:~\braket{e_0}{\Psi}=a}$, where $a=1/\sqrt{2\lambda}$. Moreover, $O$ is the origin of $\mcal{H}$, and $X$ is the point in $\mcal{H}$ such that $\overrightarrow{OX}=a\ket{e_0}$. Note that $\overrightarrow{OX}$ is perpendicular to $L_2$. The red line denotes the intersection of $L_1$ and $L_2$, and the points on this line correspond to the flows consistent with $\viext$. Let $Y$ be an arbitrary point on the red line. Then we have $\overrightarrow{OX} \perp \overrightarrow{XY}$. Furthermore, let $Z$ be the unique point on the line $\overline{OY}$ such that $\overrightarrow{XZ} \perp \overrightarrow{OY}$. Then we can show that $\|{\overrightarrow{XZ}}\|$ achieves the minimum value if and only if $Y$ corresponds to the electrical flow consistent with $\viext$, and in this case, $Z$ is exactly the projection of $X$ onto $L_1$.
} 
\label{fig:fig1}
\end{figure}

Let $\mcal{H} \defeq \spn{\ket{e}: e \in E'}$. For any $A \in \mcal{H}$, we view $A$ as both a vector and a point, and for any $A, B \in \mcal{H}$, we  define $\overrightarrow{AB}$ as the vector from the point $A$ to the point $B$. Let $O$ be the origin of $\mcal{H}$, and let $X$ be the point in $\mcal{H}$ such that $\overrightarrow{OX}=a\ket{e_0}$, where $a \defeq 1/\sqrt{2\lambda}$. Then $L_1 \defeq \ker{C_{G'}}$ is a hyperplane in $\mcal{H}$. Moreover, let $L_2 \defeq \seta{ \ket{\Psi} \in \mcal{H}:~\braket{e_0}{\Psi}=a}$. Then, for any $\ket{\psi} \in L_2$, we can write it as $\ket{\Psi}=a \ket{e_0}+\ket{\Psi'}$ for some vector $\ket{\Psi'} \bot \ket{e_0}$. So $L_2$ is a hyperplane orthogonal to the vector $\overrightarrow{OX}$ and it also touches the point $X$. 

Now consider $L_3 \defeq L_1 \cap L_2$. One can see that
\beq
L_3=\seta{g(\vf)=a \ket{e_0}+\dsum_{e \in E} \dfrac{\vf(e)}{\sqrt{w_{e}}}\ket{e}:~\vf \text{~is~a~flow~consistent~with~} \viext }.
\eeq
So the points in $L_3$ correspond to the flows consistent with $\viext$.

Let us pick arbitrary $Y \in L_3$. Then $\overrightarrow{OY}=g(\vf)$ for some flow $\vf$ consistent with $\viext$. Note that $\overrightarrow{XY}=\overrightarrow{OY}-\overrightarrow{OX}=\sum_{e \in E} \frac{\vf(e)}{\sqrt{w_e}}\ket{e}$ and hence $\nm{\overrightarrow{XY}}^2=\mcal{E}(\vf)$. Then, since $\overrightarrow{OX} \perp \overrightarrow{XY}$ and $\nm{\overrightarrow{OX}}=a$, we get $\nm{\overrightarrow{OY}}^2=a^2+\mcal{E}(\vf)$. Now let $Z$ be the unique point in the line $\overline{OY}$ that is closest to $X$. Then we have $\overrightarrow{XZ} \perp \overrightarrow{OY}$, and hence $\nm{ \overrightarrow{XZ} }= a \sqrt{\mcal{E}(\vf)/(a^2+\mcal{E}(\vf))}$. Note that $\nm{ \overrightarrow{XZ} }$ achieves the minimum value if and only if $\mcal{E}(\vf)$ achieves the minimum value. By Lemma \ref{lem:minpower}, the electrical flow $\vi$ has the minimum power among all the flows consistent with $\viext$. So $\nm{ \overrightarrow{XZ} }$ achieves the minimum value if and only if $Y$ corresponds to the electrical flow $\vi$, i.e. $\overrightarrow{OY}=g(\vi)$.
Then the corresponding $Z$ is the point closest to $X$ in $L_1$. In other words, this $Z$ is exactly the projection of $X$ onto $L_1$. So we have $\Pi \ket{e_0} =\overrightarrow{OZ} \propto \overrightarrow{OY} = g(\vi)$, as claimed.
}

\subsection{Defining quantum walks}
\label{sec:definequantumwalk}
Now we define a quantum walk \cite{Amb03, Sze04} related to the matrix $C_{G'}$, analyze its spectral properties, and give its efficient implementations. This quantum walk will become a key component of the algorithms in the next subsection. It can be viewed as a generalization of those used for evaluating span programs \cite{Rei09,Rei10,BR12b}.

Let us define two operators $A$ and $B$ as follows:
\bea
A \defeq \dsum_{v \in V} \ket{\psi_v}\ket{v}\bra{v}, \\
B \defeq \dsum_{e \in E'} \ket{e} \ket{\phi_e} \bra{e},
\eea
where
\bea
\ket{\psi_v} &\defeq & \dfrac{1}{\sqrt{\wdgp{v}}} \dsum_{e \in E'(v)} \sqrt{w_e} \ket{e}  \\
&=&
\dfrac{1}{\sqrt{\wdg{v}+\lambda}} \lb \sqrt{\lambda} \ket{e_0} + \dsum_{e \in E(v)} \sqrt{w_e} \ket{e}  \rb, ~~~~ \forall v \in V,
\label{eq:psiv}
\eea
and
\bea
\ket{\phi_e} &\defeq & \dfrac{1}{\sqrt{2}}\lb \ket{e^+}-\ket{e^-} \rb, ~~~~ \forall e \in E,\label{eq:phie0}\\
\ket{\phi_{e_0}} &\defeq & \ket{\viext} ~=
\dsum_{v \in V} {\viext(v)} \ket{v}. 
\label{eq:phie}
\eea
Note that the $\ket{\psi_v}$'s and $\ket{\phi_e}$'s are all unit vectors (recall that $\nm{\viext}=1$). So $A$ and $B$ are both isometries, and 
\bea
AA^{\dagger} &=& \proj{A},\\
BB^{\dagger} &=& \proj{B}.
\eea
Furthermore, by a direct computation, one can check that
\beba
D(A, B) \defeq A^{\dagger}B
=-\dfrac{1}{\sqrt{2}} D_{G'}^{-1/2} C_{G'}.
\eeea
This implies that $\ker{D(A, B)} = \ker{C_{G'}}$, which is characterized by Lemma \ref{lem:kercg}.

Now we define a unitary operator $U(A, B)$ as follows:
\beba
U(A, B) \defeq \refl{B}\cdot \refl{A}.
\eeea
We can find the eigenvalues and eigenvectors of $U(A, B)$ by using Szegedy's spectral lemma:

\begin{lemma}[Spectral Lemma, \cite{Sze04}]
Let $\mcal{H}$ be a Hilbert space, and let $A$, $B$ be two operators such that
$AA^{\dagger}=\proj{A}$ and $BB^{\dagger}=\proj{B}$ are both projections onto  subspaces of $\mcal{H}$. Let $D(A, B) \defeq A^{\dagger}B$ and let $U(A, B)\defeq \refl{B}\cdot \refl{A}$. Then all the singular values of $D(A, B)$ are at most $1$. Let $\{\cos{\theta_j}:~1 \le j \le k\}$ be the singular values of $D(A, B)$ that lie in the open interval $(0, 1)$ (counted with multiplicity), and let $\{\ket{w_j}, \ket{u_j}:~1 \le j \le k\}$ be the associated left and right singular vectors. Then those eigenvalues of $U(A, B)$ that have nonzero imaginary part are exactly 
\beba
\{e^{-2i\theta_j}, ~e^{2i\theta_j}:~1 \le j \le k\}.
\eeea
The (unnormalized) eigenvectors associated with these eigenvalues are 
\beba
\{A\ket{w_j}-e^{-i \theta_j} B\ket{u_j}, ~A\ket{w_j}-e^{i \theta_j} B\ket{u_j}:~1 \le j \le k\}.
\eeea
Furthermore,
\ben
\item The $+1$ eigenspace of $U(A, B)$ is $\lb \range{A} \cap \range{B}\rb \oplus \lb \range{A}^{\perp} \cap \range{B}^{\perp}\rb$. 
\item The $-1$ eigenspace of $U(A, B)$ is $\lb \range{A} \cap \range{B}^{\perp}\rb \oplus \lb \range{A}^{\perp} \cap \range{B}\rb$. In addition, $\range{A}^{\perp} \cap \range{B}=\seta{B\ket{u}:~\ket{u} \in \ker{D(A, B)}}$.
\een
The above is a complete description of the eigenvalues and eigenvectors of operator $U(A, B)$ acting on $\mcal{H}$.
\label{lem:spectrallemma}
\end{lemma}

We are interested in the $-1$ eigenspace of $U(A, B)$. Let $\mcal{H}'$ be this subspace, and let $\Pi'$ be the projection onto this subspace. By Lemma \ref{lem:spectrallemma}, $\mcal{H}'$ is the direct sum of $\mcal{H}'' \defeq \seta{B  \ket{u}:~\ket{u} \in \ker{D(A,B)} }$ and another subspace which is orthogonal to $\range{B}$. Then, since $B$ is an isometry, Lemma \ref{lem:projkercg} implies that
\beba
\Pi' B\ket{e_0} \propto B \ket{\Phi} = a B \ket{e_0} + B W_G^{-1/2} \ket{\vi} 
= a \ket{\phi_{e_0}} \ket{e_0} + \dsum_{e \in E} \dfrac{\vi(e)}{\sqrt{w_e}} \ket{\phi_e} \ket{e},
\label{eq:projbe0}
\eeea
where $a \defeq 1/\sqrt{2\lambda}$. This fact will be useful in the next subsection.

The following lemma gives a lower bound on the eigenphase gap around $\pi$ of $U(A, B)$:

\begin{lemma}
The eigenphase gap around $\pi$ of $U(A, B)$ is at least $\sqrt{2\lambda/3}$.
\label{lem:spectralgapqw}
\end{lemma}
\proof{ Since $\lambda_2(\nlap) \ge \lambda>0$, by Lemma \ref{lem:nlapgnlapgpspectralgap}, we have $\lambda_2(\nlapp)  \ge \lambda/3>0$. Then, by
\beba
D(A, B)D(A, B)^{\dagger} = \dfrac{1}{2} D_{G'}^{-1/2}C_{G'}C_{G'}^T D_{G'}^{-1/2}
=\dfrac{1}{2} \nlapp,
\eeea
we get
$s_2(D(A, B)) \ge \sqrt{\lambda/6}$. Meanwhile, by Lemma \ref{lem:spectrallemma}, the singular value $s_j \in (0, 1)$ of $D(A,B)$ is mapped to the eigenvalues  $e^{\pm 2 i \arccos{s_j}}$ of $U(A, B)$. Let $\theta_j = \pi/2 - \arccos{s_j}$. Then we have
\beba
\theta_j \ge \sin{\theta_j} = s_j \ge \sqrt{\dfrac{\lambda}{6}}.
\eeea
Therefore, the eigenphase gap around $\pi$ of $U(A, B)$ is at least $2\sqrt{\lambda/6}=\sqrt{2\lambda/3}$, as claimed.
}

The following lemmas give upper bounds on the cost of implementing $U(A, B)$ perfectly or approximately:

\begin{lemma}
$U(A,B)$ can be implemented by a gate-efficient procedure that makes $\bgo{d}$ uses of $\mcal{P}_v$, $\mcal{P}_e$ and $\mcal{P}_i$. 
\label{lem:qwresource}
\end{lemma}
\proof{ Since $U(A,B)=\refl{B} \cdot \refl{A}$, we only need to show that both $\refl{A}$ and $\refl{B}$ can be implemented by gate-efficient procedures that make $\bgo{d}$ uses of $\mcal{P}_v$, $\mcal{P}_e$ and $\mcal{P}_i$..

To implement $\refl{A}$, we use the following method. Let $Q_1$ be a unitary operation that maps $\ket{0^n} \ket{v}$ to $\ket{\psi_v}\ket{v}$ for all $v \in V$, and let $R_1$ be the reflection about $\spn{\ket{0^n}\ket{v}: ~v \in V}$, where $n=\tht{\log{N}}$. Then
\beba
\refl{A} = Q_1R_1Q_1^{\dagger}.
\eeea
Clearly, $R_1$ can be implemented in time $\poly{\log{N}}$. We implement $Q_1$ using the following procedure. Given the state $\ket{0^n}\ket{v}$ for any $v \in V$, we first map it to
\beba
\ket{0^n} \ket{v} \lb \bigotimes\limits_{e \in E(v)} \ket{e} \ket{w_e} \rb
\eeea
by using $\bgo{d}$ queries to $\mcal{P}_{v}$ and $\mcal{P}_e$ (recall that $\card{E(v)} \le d$). Then we transform this state into
\beba
\ket{\psi_v} \ket{v} \lb \bigotimes\limits_{e \in E(v)} \ket{e} \ket{w_e} \rb,
\eeea
where
\beba
\ket{\psi_v} = \dfrac{1}{\sqrt{\wdg{v}+\lambda}}\lb \sqrt{\lambda} \ket{e_0}+
\dsum_{e \in E(v)} \sqrt{w_{e}} \ket{e} \rb.
\eeea
Since $\ket{\psi_v}$ is a $(d+1)$-sparse vector in a $\poly{N}$-dimensional space, this step can be accomplished by using $O(d \cdot \poly{\log{N}})$ 2-qubit gates, as implied by Ref.~\cite{SBM04}. Finally, we uncompute $\bigotimes_{e \in E(v)} \ket{e} \ket{w_e}$ by using $\bgo{d}$ queries to $\mcal{P}_{v}$ and $\mcal{P}_e$. This implementation of $Q_1$ requires $\bgo{d}$ uses of $\mcal{P}_v$ and $\mcal{P}_e$, and is gate-efficient. As a result, $\refl{A}$ can be implemented by a gate-efficient procedure that makes $\bgo{d}$ uses of $\mcal{P}_v$ and $\mcal{P}_e$.

The implementation of $\refl{B}$ is similar. Let $Q_2$ be a unitary operation that maps $\ket{e} \ket{0^m}$ to $\ket{e}\ket{\phi_e}$ for all $e \in E'$, and let $R_2$ be the reflection about $\spn{\ket{e}\ket{0^m}: ~e \in E'}$, where $m=\tht{\log{N}}$. Then we have
\beba
\refl{B} = Q_2R_2Q_2^{\dagger}.
\eeea
Clearly, $R_2$ can be implemented in time $\poly{\log{N}}$. We implement $Q_2$ using the following procedure. Given the state $\ket{e}\ket{0^m}$ for any $e \in E'$, if $e=e_0$, then we transform $\ket{0^m}$ into $\ket{\phi_{e_0}}=\ket{\viext}$ by calling $\mcal{P}_i$ once; otherwise, we first map this state to
\beba
\ket{e}\ket{0^m} \lb \ket{e^+} \ket{e^-} \ket{w_e}\rb
\eeea
by using $\bgo{1}$ queries to $\mcal{P}_e$, then transform it into
\beba
\ket{e}\ket{\phi_e} \lb \ket{e^+} \ket{e^-} \ket{w_e}\rb,
\eeea
where $\ket{\phi_e} =( \ket{e^+}-\ket{e^-})/\sqrt{2}$, by using $\poly{\log{N}}$ 2-qubit gates, and finally uncompute $\ket{e^+} \ket{e^-} \ket{w_e}$ by using $\bgo{1}$ queries to $\mcal{P}_e$. This implementation of $Q_2$ requires $\bgo{1}$ uses of $\mcal{P}_e$ and $\mcal{P}_i$, and is gate-efficient. As a consequence, $\refl{B}$ can be implemented by a gate-efficient procedure that makes $\bgo{1}$ uses of $\mcal{P}_e$ and $\mcal{P}_i$.
}

\begin{lemma}
$U(A,B)$ can be implemented with precision $\delta>0$ by a gate-efficient procedure that makes $$\bgo{\sqrt{\dfrac{c}{\lambda}}\cdot  \poly{\log{\dfrac{d}{\delta}}}}$$ uses of $\mcal{P}_v$, $\mcal{P}_e$ and $\mcal{P}_i$.
\label{lem:qwresource2}
\end{lemma}
\proof{ Let us use the same notation as in the proof of Lemma \ref{lem:qwresource}. Recall that $R_1$, $R_2$ and $Q_2$ can be all implemented by gate-efficient procedures that make $\bgo{1}$ uses of $\mcal{P}_v$, $\mcal{P}_e$ and $\mcal{P}_i$. So we only need to show that $Q_1$ can be implemented
with precision $\delta>0$ by a gate-efficient procedure that makes $O(\sqrt{c/\lambda}\cdot  \poly{\log{d/\delta}})$ uses of $\mcal{P}_v$ and $\mcal{P}_e$.

Recall that $Q_1$ is the unitary operation mapping $\ket{0^n}\ket{v}$ to $\ket{\psi_v}\ket{v}$ for all $v \in V$, where $n=\tht{\log{N}}$ and
\beba
\ket{\psi_v} = \dfrac{1}{\sqrt{\wdgp{v}}}
\dsum_{e \in E'(v)} \sqrt{w_{e}} \ket{e}.
\eeea
Given the state $\ket{0^n}\ket{v}$ for any $v \in V$, we first map it to
\beba
\ket{0^n}\ket{v} \ket{d_v},
\eeea
where $d_v \defeq \card{E(v)} = \uwdg{v}$, by using $\bgo{\log{d}}$ queries to $\mcal{P}_v$ and $\poly{\log{N}}$ 2-qubit gates (via binary search). Then, we transform it into
\beba
\lb  \dfrac{1}{\sqrt{d_v+1}}\dsum_{j=0}^{d_v}\ket{j}  \rb \ket{v} \ket{d_v} 
\eeea
by using $\poly{\log{N}}$ 2-qubit gates. Next, we convert this state into
\beba
 \lb \dfrac{1}{\sqrt{d_v+1}}\dsum_{ e \in E'(v)}\ket{e} \rb \ket{v} \ket{d_v} 
 =
  \lbb \dfrac{1}{\sqrt{d_v+1}} \lb \ket{e_0}+\dsum_{ e \in E(v)}\ket{e} \rb \rbb \ket{v} \ket{d_v} 
\eeea
by using $\bgo{1}$ queries to $\mcal{P}_v$ and $\poly{\log{N}}$ 2-qubit gates. Then, we transform this state into
\beba
\lb \dfrac{1}{\sqrt{d_v+1}} \dsum_{ e \in E'(v)}\ket{e} \ket{w_e} \rb \ket{v} \ket{d_v}
=
\lbb \dfrac{1}{\sqrt{d_v+1}} \lb \ket{e_0} \ket{\lambda} + \dsum_{ e \in E(v)}\ket{e} \ket{w_e} \rb \rbb \ket{v} \ket{d_v}
\eeea
by using $\bgo{1}$ queries to $\mcal{P}_e$ and $\poly{\log{N}}$ 2-qubit gates. After that, we append an ancilla qubit in state $\ket{0}$, and perform the following controlled-rotation:
\beba
\ket{w_e}\ket{0} \to
\ket{w_e}\lb \sqrt{\dfrac{w_e}{2c}}\ket{0}+ \sqrt{1-\dfrac{w_e}{2c}}\ket{1} \rb.
\eeea
This is a valid unitary operation, because $w_e \le 2c$ for all $e \in E'(v)$ (note that $w_{e_0}=\lambda \le 2 \le 2c$). Then, we measure the ancilla qubit, and conditioning on the outcome being $0$, we obtain the state 
\beba
\lb \dfrac{\sum_{e \in E'(v)} \sqrt{w_e} \ket{e} \ket{w_e}}{\nm{\sum_{e \in E'(v)} \sqrt{w_e} \ket{e} \ket{w_e}}} \rb \ket{v} \ket{d_v}.
\eeea
The probability of this event happening is $\omg{\lambda/c}$, since $w_e \ge \lambda/2$ for all $e \in E'(v)$ (note that $w_e \ge 1 \ge \lambda/2$ for all $e \in E$). Next, we uncompute $\ket{d_v}$ by using $\bgo{\log{d}}$ queries to $\mcal{P}_v$ and $\poly{\log{N}}$ 2-qubit gates. Finally, we uncompute $\ket{w_e}$ by using $\bgo{1}$ queries to $\mcal{P}_e$ and $\poly{\log{N}}$ 2-qubit gates, and obtain the desired state $ \ket{\psi_v} \ket{v}$. 

The above procedure, denoted by $\mcal{A}$, makes $\bgo{\log{d}}$ uses of $\mcal{P}_v$ and $\mcal{P}_e$, is gate-efficient, and 
has $\omg{\lambda/c}$ success probability. We can raise the success probability to $\omg{1}$ by using the standard amplitude amplification, which requires $\bgo{\sqrt{c/\lambda}}$ repetitions of $\mcal{A}$. Let $\mcal{A}'$ be this modified procedure with $\omg{1}$ success probability. Then we can further boost the success probability to $1-\bgo{\delta'}$ by using Grover's $\pi/3$ amplitude amplification (i.e. the generalization of fixed-point quantum search) \cite{Gro05}, which requires $\bgo{\log{1/\delta'}}$ repetitions of $\mcal{A}'$. Let us pick $\delta'=\tht{\delta^2}$, and let $\mcal{A}''$ be this procedure with $1-\bgo{\delta^2}$ success probability. Then $\mcal{A}''$ makes $\bgo{\sqrt{c/\lambda} \cdot \poly{\log{d/\delta}}}$ uses of $\mcal{P}_v$ and $\mcal{P}_e$, is gate-efficient, and satisfies
\beba
\mcal{A}'' \ket{0^t} \ket{0^n} \ket{v} 
= \sqrt{1-\delta_v} \ket{0^t} \ket{\psi_v} \ket{v} 
+  \ket{\Phi_v^{\perp}},
\label{eq:qwapp}
\eeea
where $t$ is a positive integer, $\delta_v =\bgo{\delta^2}$,
$\ket{\Phi_v^{\perp}}$ is an unnormalized state satisfying 
 $\braket{\Phi_v^{\perp}}{\Phi_v^{\perp}}=\delta_v$ and
$(\ketbra{0^t}{0^t} \otimes I) \ket{\Phi_v^{\perp}}=0$, for all $v \in V$. This implies that
\beba
\nm{ \mcal{A}'' \ket{0^t}  \ket{0^n} \ket{v}
- \ket{0^t}  \ket{\psi_v} \ket{v} }^2
=(1- \sqrt{1-\delta_v})^2+\delta_v
=\bgo{\delta^2}.
\eeea
Meanwhile, since $\mcal{A}''$ is a unitary operation, we have $\mcal{A}'' \ket{0^t} \ket{0^n} \ket{u} \perp \mcal{A}'' \ket{0^t} \ket{0^n} \ket{v}$ for any $u \neq v$. Then by Eq.~(\ref{eq:qwapp}), we know that $\ket{0^t}\ket{\psi_u}\ket{u}$, $\ket{0^t}\ket{\psi_v}\ket{v}$, $\ket{\Phi_u^{\perp}}$ and $\ket{\Phi_v^{\perp}}$ are mutually orthogonal for any $u \neq v$. As a result, for any normalized state $\ket{z}=\sum_{v \in V} z_v \ket{v}$, we have
\bea
\nm{ \mcal{A}'' \ket{0^t}  \ket{0^n} \ket{z}
- \ket{0^t} (Q_1 \ket{0^n} \ket{z} )}^2
&=&\nm{\dsum_{v \in V} z_v \lb\mcal{A}'' \ket{0^t} \ket{0^n} \ket{v} 
-\ket{0^t}  \ket{\psi_v} \ket{v} \rb}^2 \\
&=& \dsum_{v \in V} \abs{z_v}^2 \nm{\mcal{A}'' \ket{0^t} \ket{0^n} \ket{v} 
-\ket{0^t} \ket{\psi_v} \ket{v}}^2 \\
&=&\bgo{\delta^2}.
\eea
This means that 
\beba
\nm{\bra{0^t}\mcal{A}'' \ket{0^t} - Q_1}=\bgo{\delta},
\eeea
as desired.
}

\subsection{Using quantum walks to analyze electrical networks}
\label{sec:aenbyqw}
Now we describe our quantum-walk-based algorithms for solving the ENA-P and ENA-ER problems. These algorithms require the following variant of phase estimation \cite{Kit95,CEMM97}, which determines whether the eigenphase corresponding to an eigenvector of a unitary operation is $\theta$ or far away from $\theta$, for some given $\theta \in [0, 2\pi)$, succeeding with probability close to $1$. (Similar procedures have been used in e.g. Refs.~\cite{NWZ09, CKS15}.)

\begin{lemma}
Let $U$ be a unitary operation with eigenvectors $\ket{\psi_j}$ satisfying
$U \ket{\psi_j} = e^{i \theta_j} \ket{\psi_j}$ for some $\theta_j \in [0, 2\pi)$. Let $\theta \in [0, 2\pi)$ and let $\Delta, \delta \in (0, 1)$. Then there is a unitary procedure $\mcal{P}$ that requires $\bgo{(1/\Delta)\cdot \log{1/\delta}}$ uses of $U$ and $\poly{\log{1/(\Delta\delta)}}$ additional 2-qubit gates, and satisfies
\beba
\mcal{P} \ket{0}\ket{0^l} \ket{\psi_j} = \lb \alpha_{j,0} \ket{0} \ket{\eta_{j, 0}} + \alpha_{j, 1} \ket{1} \ket{\eta_{j, 1}} \rb \ket{\psi_j},
\label{eq:p00lpsi}
\eeea
where $l=\bgo{\log{1/\Delta} \log{1/\delta}}$, $\abs{\alpha_{j, 0}}^2 + \abs{\alpha_{j, 1}}^2=1$, $\ket{\eta_{j, 0}}$ and $\ket{\eta_{j, 1}}$ are two normalized states, and
\bit
\item If $\theta_j=\theta$, then $\abs{\alpha_{j, 0}}^2 \ge 1-\delta$.
\item If $\abs{\theta_j-\theta} \ge \Delta$, then $\abs{\alpha_{j, 1}}^2 \ge 1-\delta$.
\eit
\label{lem:boostedpe}
\end{lemma}
\proof{ We can get a $\Delta/2$-additive approximation of $\theta_j$ by using the standard phase estimation, which requires $\bgo{1/\Delta}$ uses of $U$ and $\poly{\log{1/\Delta}}$ additional 2-qubit gates. This is sufficient to distinguish between the two cases. However, it only succeeds with $\omg{1}$ probability. To overcome this issue, we repeat this procedure $\bgo{\log{1/\delta}}$ times and check whether the median of the estimates is $\Delta/2$-close to $\theta$. By a standard Chernoff bound, we can ensure that the failure proability is at most $\delta$. Let $\mcal{P}$ be this boosted procedure. Then $\mcal{P}$ requires $\bgo{(1/\Delta) \cdot \log{1/\delta}}$ uses of $U$ and $\poly{\log{1/(\Delta\delta)}}$ additional 2-qubit gates, and satisfies the desired properties.
}

\begin{theorem}
The ENA-P problem can be solved by a gate-efficient quantum algorithm that makes 
$$\bgo{\mmin{\dfrac{c^{0.5}d^{1.5}}{\epsilon \lambda}, ~\dfrac{c d^{0.5}}{\epsilon \lambda^{1.5}}} \cdot \poly{\log{\dfrac{cd}{\epsilon \lambda}}}}$$ uses of $\mcal{P}_v$, $\mcal{P}_e$ and $\mcal{P}_i$.
\label{thm:enap2}
\end{theorem} 
\proof{ \noindent\textbf{Algorithm:}
We estimate $\mcal{E}(\vi)$ up to multiplicative error $\bgo{\epsilon}$ by using the following algorithm:
\bit
\item Let $\mcal{P}$ be the unitary procedure in Lemma \ref{lem:boostedpe} for $U=U(A,B)$, $\theta=\pi$, $\Delta= \sqrt{\lambda/3}$ and $\delta=\bgo{\epsilon \lambda/(cd)}$. Suppose
\beba
\mcal{P}\ket{0}_1\ket{0^l}_2 (B\ket{e_0})_3
=\mu_0 \ket{0}_1 \ket{\vphi_0}_{2,3} 
+\mu_1 \ket{1}_1 \ket{\vphi_1}_{2,3} 
\eeea
where $l=\bgo{\log{1/\Delta} \log{1/\delta}}$, $\abs{\mu_0}^2+\abs{\mu_1}^2 = 1$, $\ket{\vphi_0}$ and $\ket{\vphi_1}$ are two normalized states. We use amplitude estimation to get an $\bgo{\epsilon}$-multiplicative approximation $\hat{r}$ of $r \defeq \abs{\mu_1}^2$ (succeeding with probability at least $3/4$). Then we return 
\beba 
\hat{E} \defeq \dfrac{ \hat{r}}{1- \hat{r}}\cdot \dfrac{1}{ 2\lambda}
\label{eq:hatE}
\eeea
as our estimate of $\mcal{E}(\vi)$. During this process, the unitary operation $U(A, B)$ is implemented either by the procedure in Lemma \ref{lem:qwresource}, or by the procedure in Lemma \ref{lem:qwresource2} with precision $\bgo{ \epsilon^{2} \lambda^{2}  / (c d)}$.\\
\eit

\noindent\textbf{Correctness:}
Recall that $\mcal{H}'$ is the $-1$ eigenspace of $U(A, B)$, and $\Pi'$ is the projection onto this subspace. We have shown in the previous subsection that
\bea
\Pi' B \ket{e_0} \propto  B \ket{\Phi} = a B \ket{e_0}	 + B W^{-1/2} \ket{\vi},
\label{eq:pipbe0}
\eea
where $a = 1/{\sqrt{2\lambda}}$. Since $B$ is an isometry, we have
\bea
\nm{BW^{-1/2}\ket{\vi}}^2
=\nm{W^{-1/2}\ket{\vi}}^2
= \mcal{E}(\vi) = \viext^T \lap^+ \viext.
\eea
Meanwhile, since $\lambda_2(\nlap) \ge \lambda$ and $1 \le \wdg{v} \le cd$ for all $v \in V$, by Eq.~(\ref{eq:eigvallap}) and Lemma \ref{lem:lapnlapspectralgap}, we get
\bea
\lambda \le \lambda_2(\lap) \le \lambda_3(\lap) \le \dots \le \lambda_N(\lap) \le 2cd.
\eea
Then, since $\viext \in \range{\lap}$ and $\nm{\viext}=1$, we obtain
\bea
\dfrac{1}{2cd} \le \mcal{E}(\vi) = \viext^T \lap^+ \viext \le \dfrac{1}{\lambda}=2a^2.
\label{eq:evibound}
\eea

Now let
\bea
\ket{\Psi} \defeq \dfrac{B\ket{\Phi}}{\nm{B\ket{\Phi}}}
=\dfrac{a B \ket{e_0}	 + B W^{-1/2} \ket{\vi}}{\sqrt{a^2+\mcal{E}(\vi)}} \in \mcal{H}',
\eea
and let $\seta{\ket{\Psi_k^{\perp}}:~1\le k \le K}$ be an orthonormal basis for $(\mcal{H}')^{\perp}$. Then Eq.~(\ref{eq:pipbe0}) implies
\bea
B\ket{e_0} = \beta \ket{\Psi} + \dsum_{k=1}^K \beta_k \ket{\Psi_k^{\perp}}, 
\label{eq:beodecomp}
\eea
for some numbers $\beta$, $\beta_1, \beta_2, \dots, \beta_K$. Since $B$ is an isometry, we have
\beba
\beta = \bra{\Psi}B{\ket{e_0}}=\dfrac{a}{\sqrt{a^2+\mcal{E}(\vi)}}.
\eeea
Let
\beba
r_1 \defeq 1- \abs{\beta}^2
=\dsum_{k=1}^K \abs{\beta_k}^2 = \dfrac{\mcal{E}(\vi)}{a^2+\mcal{E}(\vi)}.
\label{eq:defr1}
\eeea
Then we have
\beba
\mcal{E}(\vi) = \dfrac{r_1}{1-r_1} \cdot a^2.
\label{eq:evir1relation}
\eeea
In addition, by Eqs.~(\ref{eq:evibound}) and (\ref{eq:defr1}), we get
\bea
\dfrac{1}{\kappa+1} \le r_1 \le \dfrac{2}{3},
\label{eq:r1bounds}
\eea
where $\kappa \defeq cd/\lambda$. 

Now, since $\Delta=\sqrt{\lambda/3}$ is smaller than the eigenphase gap around $\pi$ of $U(A, B)$ by Lemma \ref{lem:spectralgapqw}, $\mcal{P}$ satisfies
\beba
\mcal{P} \ket{0} \ket{0^l} \ket{\Psi}
=\lb \alpha_0 \ket{0} \ket{\eta_0} + \alpha_1 \ket{1} \ket{\eta_1}\rb \ket{\Psi},
\eeea
where $\abs{\alpha_0}^2 \ge 1-\delta$, $\abs{\alpha_1}^2 \le \delta$, $\ket{\eta_0}$ and $\ket{\eta_1}$ are normalized states, and
\beba
\mcal{P} \ket{0} \ket{0^l} \ket{\Psi_k^{\perp}}
=\lb \alpha_{k, 0} \ket{0} \ket{\eta_{k,0}} + \alpha_{k, 1} \ket{1} \ket{\eta_{k,1}}  \rb \ket{\Psi_k^{\perp}},
\eeea
where $\abs{\alpha_{k, 1}}^2 \ge 1-\delta$, $\abs{\alpha_{k, 0}}^2 \le \delta$, 
 $\ket{\eta_{k, 0}}$ and $\ket{\eta_{k, 1}}$ are normalized states, for all $k$. As a result, we get
\bea
\mcal{P} \ket{0} \ket{0^{l}} (B\ket{e_0})
&=& \ket{0}\lb \alpha_0 \beta  \ket{\eta_0} \ket{\Psi}
+ \dsum_{k=1}^K \alpha_{k, 0} \beta_k \ket{\eta_{k, 0}} \ket{\Psi_k^{\perp}}\rb\\
&+&\ket{1}\lb \alpha_1 \beta \ket{\eta_1} \ket{\Psi}
+ \dsum_{k=1}^K \alpha_{k, 1} \beta_k  \ket{\eta_{k, 1}} \ket{\Psi_k^{\perp}}\rb.
\eea
This implies that
\beba
r = \abs{\alpha_1 \beta}^2 + \dsum_{k=1}^K \abs{\alpha_{k, 1} \beta_k}^2. 
\eeea
Note that 
\beba
\abs{r_1-\dsum_{k=1}^K \abs{ \alpha_{k, 1} \beta_k }^2}
=\dsum_{k=1}^K (1-\abs{\alpha_{k,1}}^2)\abs{\beta_k}^2
\le \delta r_1 = \bgo{\epsilon r_1},
\eeea
and
\beba
\abs{\alpha_1 \beta}^2 \le \abs{\alpha_1}^2 \le \delta =\bgo{\epsilon r_1},
\eeea
since $r_1=\omg{1/\kappa}=\omg{\lambda/(cd)}$ by Eq.~(\ref{eq:r1bounds}). It follows that
\beba
\abs{r - r_1} \le 
\abs{r_1 - \dsum_{k=1}^K \abs{ \alpha_{k, 1} \beta_k }^2}
+\abs{\alpha_1 \beta}^2 
=\bgo{\epsilon r_1}.
\eeea
Namely, $r$ is an $\bgo{\epsilon}$-multiplicative approximation of $r_1$. Meanwhile, $\hat{r}$ is an $\bgo{\epsilon}$-multiplicative approximation of $r$. 
Combining these two facts, we know that
\beba
\abs{\hat{r}-r_1} 
\le \abs{\hat{r} - r} + \abs{r-r_1}
=\bgo{\epsilon r} + \bgo{\epsilon r_1}
=\bgo{\epsilon r_1},
\label{eq:hatrr1}
\eeea
and 
\beba
\abs{(1-\hat{r}) -(1-r_1 )}=\abs{\hat{r}-r_1} = \bgo{\epsilon r_1} =\bgo{\epsilon (1-r_1)},
\label{eq:hatrr2}
\eeea
since $r_1 \le 2/3$ by Eq.~(\ref{eq:r1bounds}). This implies that
\beba
(1-\bgo{\epsilon}) \cdot \dfrac{r_1}{1-r_1}
\le \dfrac{\hat{r}}{1-\hat{r}}
\le (1+\bgo{\epsilon}) \cdot \dfrac{r_1}{1-r_1},
\eeea
and hence
\beba
\abs{\dfrac{\hat{r}}{1-\hat{r}}-\dfrac{r_1}{1-r_1}} = \bgo{\epsilon \cdot \dfrac{r_1}{1-r_1}}. 
\eeea
As a consequence, by Eqs.~(\ref{eq:hatE}) and (\ref{eq:evir1relation}), we get
\beba
\abs{\hat{E}-\mcal{E}(i)}
=\abs{\dfrac{\hat{r}}{1-\hat{r}}\cdot a^2-\dfrac{r_1}{1-r_1} \cdot a^2}
=\bgo{\epsilon \cdot \dfrac{r_1}{1-r_1} \cdot a^2}
=\bgo{\epsilon \cdot \mcal{E}(\vi)},
\eeea
as desired.

In the above argument, we have assumed that the unitary operation $U(A, B)$ is implemented perfectly. This is true if we use the procedure in Lemma \ref{lem:qwresource} to implement $U(A, B)$. If we instead use the procedure in Lemma \ref{lem:qwresource2} to implement $U(A, B)$ with precision $\bgo{\lambda^2 \epsilon^2/(cd)}$, then the algorithm still outputs a correct $\hat{E}$ with high probability. The reason is as follows. We will show below that this algorithm only makes $\lto{cd/(\lambda^2 \epsilon^2)}$ uses of $U(A, B)$. Provided that each $U(A, B)$ is implemented with precision $\bgo{\lambda^2 \epsilon^2/(cd)}$, the error in the final state (compared to the ideal case) is only $\lto{1}$. Therefore, the probability that this algorithm outputs a correct $\hat{r}$ (and hence a correct $\hat{E}$) is at least $3/4-\lto{1}$.\\

\noindent\textbf{Complexity:} The state $B\ket{e_0}= \ket{e_0} \ket{\viext}$ can be prepared by making $\bgo{1}$ uses of $\mcal{P}_i$. Since $r=\tht{r_1}=\omg{1/\kappa}$ and we want to estimate it up to multiplicative error $\bgo{\epsilon}$, amplitude estimation requires 
\beba
\bgo{\dfrac{1}{\epsilon\sqrt{r}}}=\bgo{\dfrac{\sqrt{\kappa}}{\epsilon}}
\eeea
repetitions of $\mcal{P}$. By Lemma \ref{lem:boostedpe}, the procedure $\mcal{P} $ can be implemented with $\bgo{(1/\Delta) \cdot \log{1/\delta}}$ uses of $U(A, B)$ and $\poly{\log{1/(\Delta \delta)}}$ additional 2-qubit gates. So this algorithm makes 
\beba
\bgo{\dfrac{\sqrt{\kappa}}{\epsilon} \cdot \dfrac{1}{\Delta} \cdot \log{\dfrac{1}{\delta}}}
=\bgo{\dfrac{\sqrt{cd}}{\epsilon \lambda} \cdot \log{\dfrac{cd}{\epsilon \lambda}}}
\eeea
uses of $U(A, B)$. If we use the procedure in Lemma \ref{lem:qwresource} to  implement $U(A, B)$, the resulting algorithm will require 
$$\bgo{\dfrac{c^{0.5} d^{1.5}}{\epsilon \lambda} \cdot \poly{\log{\dfrac{cd}{\epsilon\lambda}}}}$$ uses of $\mcal{P}_v$, $\mcal{P}_e$, $\mcal{P}_i$, and is gate-efficient. Alternatively, if we use the procedure in Lemma \ref{lem:qwresource2} to implement $U(A, B)$ with precision $O(\lambda^2 \epsilon^2/(cd))$, the resulting algorithm will require 
$$\bgo{\dfrac{c d^{0.5}}{\epsilon \lambda^{1.5}} \cdot \poly{\log{\dfrac{cd}{\epsilon \lambda}}}}$$ uses of $\mcal{P}_v$, $\mcal{P}_e$, $\mcal{P}_i$, and 
is also gate-efficient. Our claim follows from the combination of these two facts.
}

\begin{corollary}
The ENA-ER problem can be solved by a gate-efficient quantum algorithm that makes $$\bgo{\mmin{\dfrac{c^{0.5}d^{1.5}}{\epsilon \lambda}, ~\dfrac{c d^{0.5}}{\epsilon \lambda^{1.5}}} \cdot \poly{\log{\dfrac{cd}{\epsilon \lambda}}}}$$ uses of $\mcal{P}_v$ and $\mcal{P}_e$.
\label{cor:enaer2}
\end{corollary}
\proof{ Recall that $\efr(s, t) = \mcal{E}(\mbf{i})$, where $\mbf{i}=W_GB_G^T\lap^+ \chi_{s,t}$ is the electrical flow induced by the external current $\chi_{s,t}=\ket{s}-\ket{t}$. Clearly, we can prepare the state $(\ket{s}-\ket{t})/\sqrt{2}$ in time $\poly{\log{N}}$. Then we can run the algorithm in Theorem \ref{thm:enap2} to obtain an $\bgo{\epsilon}$-multiplicative approximation of $\mcal{E}(\tilde{\vi})=\mcal{E}(\vi)/2$, where $\tilde{\vi}=\vi/\sqrt{2}$ is the electrical flow induced by the external current $\chi_{s,t}/\sqrt{2}$. Then we multiply this result by a factor of $2$, and obtain an $\bgo{\epsilon}$-multiplicative approximation of $\mcal{E}(\vi)=\efr(s, t)$. By Theorem \ref{thm:enap2}, this algorithm makes 
$$\bgo{\mmin{\dfrac{c^{0.5}d^{1.5}}{\epsilon \lambda}, ~\dfrac{c d^{0.5}}{\epsilon \lambda^{1.5}}} \cdot \poly{\log{\dfrac{cd}{\epsilon \lambda}}}}$$
uses of $\mcal{P}_v$ and $\mcal{P}_e$, and is gate-efficient.
}

One can compare the algorithm in Theorem \ref{thm:enap2} (or Corollary \ref{cor:enaer2}) with the one in Theorem \ref{thm:enap1} (or Corollary \ref{cor:enaer1}). The quantum-walk-based one is unconditionally better if we use the procedure in Lemma \ref{lem:qwresource} to implement $U(A, B)$. If we instead use the procedure in Lemma \ref{lem:qwresource2} to implement $U(A, B)$, then the quantum-walk-based one has much better dependence on $d$, but slightly worse dependence on $1/\lambda$. So it is more suitable in the case where $d$ is larger than $1/\lambda$ (which is possible and common).

We remark that the algorithm in Theorem \ref{thm:enap2} can be modified to solve the ENA-C problem (and the ENA-V problem under the promise that $s$ and $t$ are adjacent vertices). Specifically, recall that
\beba
\ket{\Psi} = \dfrac{B \ket{\Phi}}{\nm{B \ket{\Phi}}}
=\dfrac{1}{\sqrt{a^2+\mcal{E}(i)}} \lb a B\ket{e_0}
+\dsum_{e \in E} \dfrac{\vi(e)}{\sqrt{w_e}} B\ket{e}
\rb.
\eeea
If we can create this state, then we can infer $\abs{\vi(e)}$ from it, for any given $e \in E$. To prepare the state $\ket{\Psi}$, we need to use a clean version of the procedure $\mcal{P}$ in Lemma \ref{lem:boostedpe}. That is, we need to replace the states $\ket{\eta_{j,0}}$ and $\ket{\eta_{j, 1}}$ in Lemma \ref{lem:boostedpe} with $\ket{0^l}$ (namely, we want to reset the $l$ ancilla qubits to their initial states after the computation). This can be approximately achieved by using the standard ``do-copy-undo" trick. Then, when we apply this clean version of $\mcal{P}$ on $\ket{0}\ket{0^{t}}B \ket{e_0}$ (for some positive integer $t$) and measure the first qubit, conditioning on the outcome being $0$, we would obtain a state close to $\ket{\Psi}$, from which $\abs{\vi(e)}$ can be learned. However, this algorithm for solving ENA-C is not more efficient than the one in Theorem \ref{thm:enac1}. So we will not present it in detail here.

\section{Lower Bounds on the Complexity of Electrical Network Analysis}
\label{sec:lowerbound}
So far we have presented two classes of quantum algorithms for analyzing electrical networks. All of these algorithms have complexities polynomial in $1/\lambda$ (and other parameters), where $\lambda$ is the spectral gap of the normalized Laplacian of the network. In this section, we show that this polynomial dependence on $1/\lambda$ is necessary. Specifically, we prove that in order to solve any of the ENA-V, ENA-C, ENA-P, ENA-ER problems, one has to make $\omg{1/\sqrt{\lambda}}$ queries to the graph. This lower bound implies that our algorithms are optimal up to polynomial factors \footnote{The polynomial dependence on the other parameters, including $c$, $d$, $\log{N}$ and $1/\epsilon$, is clearly necessary.} ~and hence cannot be greatly improved. 

\begin{theorem}
For any positive integer $N$, there exists an unweighted connected graph $G=(V, E)$ with four distinguished vertices $s, t, u, v \in V$ such that $|V|=10N$, $\uwdg{G}=3$, $(u, v)\in E$, $\lambda_2(\nlap)=\omg{1/N^2}$, $\lambda_2(\nlap)=\bgo{1/N}$, and assuming a unit electric current is injected at $s$ and extracted at $t$, one needs to make $\omg{N}$ queries to $G$ to solve any of the following problems (succeeding with probability at least $2/3$):
\ben
\item Estimate the voltage between $u$ and $v$ up to additive error $0.1$.
\item Estimate the current on $(u, v)$ up to additive error $0.1$.
\item Estimate the power dissipated by $G$ up to multiplicative error $0.1$.
\item Estimate the effective resistance between $s$ and $t$ up to multiplicative error $0.1$.
\een
\label{thm:lowerbound}
\end{theorem} 
\proof{ We will build a graph such that, if one solves any of the above problems on this graph, then one has solved a corresponding $\parity$ problem. Recall that in the $\parity$ problem, one is given oracle access to an $N$-bit string $x=x_1x_2\dots x_N$, and needs to determine the value of $\parity(x) \defeq x_1 \xor x_2 \xor \dots \xor x_N$. Our claims will follow from a known lower bound on the quantum query complexity of $\parity$.

Now let us make this argument precise. Given an $N$-bit string $x=x_1x_2 \dots x_N$, we will map it to an unweighted graph $G(x)=(V(x), E(x))$ with $10N$ vertices. For convenience, we will label the vertices in this graph by $(i, j)$ or $(i^*, j)$ for some integers $i$ and $j$. We start with $10N$ isolated vertices, which are labeled by $(i, a)$ for $i \in \intset{N+1}$ and $a \in \{0,1\}$, and $(j^*, b)$ for $j \in \intset{4N-1}$ and $b \in \{0,1\}$. Then we add the following edges to this graph:
\bit
\item For $i \in \intset{N}$ and $a \in \zo$, we add an edge between $(i, a)$ and $(i+1, a \xor x_i)$. That is, if $x_i=0$, we add an edge between $(i, 0)$ and $(i+1, 0)$, and an edge between $(i, 1)$ and $(i+1, 1)$; otherwise, we add an edge between $(i, 0)$ and $(i+1, 1)$, and an edge between $(i, 1)$ and $(i+1, 0))$.
\item For $j \in \intset{4N-2}$ and $b \in \zo$, we add an edge between $(j^*, b)$ and $((j+1)^*, b)$. 
\item For $b \in \zo$, we add an edge between $(1, 0)$ and $(1^*, b)$.
\item For $b \in \zo$, we add an edge between $((4N-1)^*, b)$ and $(N+1, b)$.
\eit
For example, Fig.\ref{fig:lowerbound} shows the graph $G(x)$ for the string $x=11010$ (where $N=5$). Note that $G(x)$ consists of $N$ crossing-type or parallel-type gadgets (where the $i$-th gadget's type depends the value of $x_i$) and two long paths, one connecting $(1,0)$ and $(N+1,0)$ and the other connecting $(1,0)$ and $(N+1,1)$. (Similar constructions have been used to prove lower bounds on the quantum query complexity of Hamiltonian simulation \cite{BACS05,BCC+13}.)

\begin{figure}[H]
\begin{center}
 \input{lower_bound_fig.tex}
\end{center}
\fcaption{The graph $G(x)$ for the string $x=11010$.} 
\label{fig:lowerbound}
\end{figure}

Now we pick $s =(1, 0)$, $t = (N+1, 0)$, $u=((2N-1)^*, 0)$ and $v=((2N)^*, 0)$. Then the graph $G(x)$ satisfies the following property: 
\bit
\item If $\parity(x)=0$, then there are two paths between $s$ and $t$ in $G(x)$. One of them is
\beba
s=(1,0) \to (2,x_1) \to (3,x_1 \xor x_2) \to \dots \to (N+1,\xor_{i=1}^N x_i)=(N+1,0)=t,
\label{eq:shortpath}
\eeea
and the other is
\beba
s=(1,0) \to (1^*, 0) \to (2^*, 0) \to \dots \to ((4N-1)^*, 0) \to (N+1, 0)=t.
\label{eq:longpath}
\eeea
\item If $\parity(x)=1$, then there is only one path between $s$ and $t$ in $G(x)$, which is described by Eq.~(\ref{eq:longpath}).
\eit

This implies that when a unit electric current is injected at $s$ and extracted at $t$, we have:
\bit
\item If $\parity(x)=0$, then there is an electrical flow of value $0.8$ on the path described by Eq.~(\ref{eq:shortpath}), a flow of value $0.2$ on the path described by Eq.~(\ref{eq:longpath}), and no flow on other edges. Thus, the current on $(u, v)$ is $0.2$, and so is the voltage between $u$ and $v$. Moreover, the power dissipated by $G(x)$ is $0.8^2\times N+0.2^2 \times 4N = 0.8 N$, and so is the effective resistance between $s$ and $t$.

\item If $\parity(x)=1$, then there is an electrical flow of value $1$ on the path described by Eq.~(\ref{eq:longpath}), and no flow on other edges. Thus, the current on $(u, v)$ is $1$, and so is the voltage between $u$ and $v$. Moreover, the power dissipated by $G(x)$ is $4N$, and so is the effective resistance between $s$ and $t$.
\eit

It follows that we can distinguish these two cases by solving any of the following problems:
\ben
\item Estimate the voltage between $u$ and $v$ up to additive error $0.1$.
\item Estimate the current on $(u, v)$ up to additive error $0.1$.
\item Estimate the power dissipated by $G$ up to multiplicative error $0.1$.
\item Estimate the effective resistance between $s$ and $t$ up to multiplicative error $0.1$.
\een 
It is known that $\parity$ has $\tht{N}$ bounded-error quantum query complexity \cite{BBC+98, FGGS98}. This implies that one needs to make $\omg{N}$ queries to $G(x)$ to solve any of the above problems.

Finally, we show that $G(x)$ satisfies the other desired properties. Clearly, $G(x)$ is a connected graph with maximum degree $3$. Moreover, it has conductance $\phi_{G(x)}=\tht{1/N}$. To see this, consider the cut $(S,\bar{S})$, where $S = \{(i,a):~ i \in \intset{\floor{N/2}}, a\in \zo\} \cup \{(j^*,b):~ j \in \intset{2N}, b \in \zo\}$ and $\bar{S} = V \setminus S$. We have $\mrm{vol}(S)$, $\mrm{vol}(\bar{S})=\tht{N}$ and $|E(S,\bar{S})|=\bgo{1}$. So $\phi_S=\bgo{1/N}$, which implies $\phi_{G(x)} = \bgo{1/N}$. On the other hand, since $G(x)$ is a connected graph with $\bgo{N}$ edges, for any cut $(S, \bar{S})$, we have $\mrm{vol}(S)$, $\mrm{vol}(\bar{S})=\bgo{N}$, $|E(S,\bar{S})|=\omg{1}$, and hence $\phi_S=\omg{1/N}$. This implies $\phi_{G(x)}=\omg{1/N}$. Combining these two facts, we obtain $\phi_{G(x)}=\tht{1/N}$. Then by Cheeger's inequality (i.e. Eq.~(\ref{eq:cheeger})), we have $\lambda_2(\overline{L}_{G(x)})=\omg{1/N^2}$ and $\lambda_2(\overline{L}_{G(x)})=\bgo{1/N}$. This concludes the proof.
}

\section{Discussion}
\label{sec:discussion}
To summarize, we have proposed two classes of quantum algorithms for analyzing large sparse electrical networks. The first class is based on solving linear systems, and the second class is based on using quantum walks. These algorithms compute various electrical quantities, including voltages, currents, dissipated powers and effective resistances, in time $\poly{d, c, \log{N}, 1/\lambda, 1/\epsilon}$, where $N$ is the number of vertices in the graph, $d$ is the maximum unweighted degree of the vertices, $c$ is the ratio of largest to smallest edge resistance, $\lambda$ is the spectral gap of the normalized Laplacian of the graph, and $\epsilon$ is the accuracy. Furthermore, we prove that the polynomial dependence on $1/\lambda$ is necessary. Hence, our algorithms are optimal up to polynomial factors and cannot be significantly improved. 

We have seen that a Laplacian system $Lx=b$ naturally arises when one wants to compute the voltages in an electrical network. Such systems also play an important role in other graph problems, such as graph partitioning (e.g.~\cite{MVM11,LGT11,MOV12,Vis12}), graph sparsification (e.g.~\cite{SS08,ST08,KLP12}) and maximum flows (e.g.~\cite{CKM+10,LRS13,Mad13}). As a result, much effort has been dedicated to studying the classical complexity of solving these systems (e.g.~\cite{ST04, ST06}). It appears  that Laplacian systems are easier to solve than general linear systems  classically. In contrast, we do not know whether the quantum analogue of this statement is true. In particular, Harrow, Hassidim and Lloyd \cite{HHL08} showed that it is $\BQP$-complete to solve a general sparse well-conditioned linear system $Ax=b$ (in certain sense). Does this theorem still hold under the restriction that $A$ is a Laplacian? If so, there would be an interesting implication: Any problem in $\BQP$ can be reduced to the problem of computing certain voltages in an exponentially-large electrical network! This can be even viewed as a novel (but impractical) proposal for building a quantum computer! On the other hand, if Laplacian systems are indeed easier to solve than general linear systems quantumly, then what is the exact quantum complexity of solving them? In particular, can they be solved in time sublinear in the Laplacian's finite condition number? These are left as interesting open questions.

In this paper, we have focused on direct-current (DC) circuits which consist of resistors and DC sources. It is also worth exploring alternating-current (AC) circuits which consist of resistors, capacitors, inductors and AC sources. Such electrical systems are governed by a set of second-order linear ordinary differential equations. But it is possible to transform these differential equations into a system of linear equations by applying the Fourier or Laplace transform. The resulting linear system can be viewed as a \emph{complex} Laplacian system, and it can be solved by invoking a quantum linear system algorithm. However, the complexity of this algorithm is difficult to analyze, because it depends on the condition number of a complex Laplacian, and there are few known methods to bound this quantity (to our knowledge, there is no analogue of Cheeger's inequality for complex Laplacians). So it is unclear how much quantum advantage can be gained on such systems.

Spectral graph theory has become a powerful tool for the design and  analysis of fast classical algorithms for various graph problems, such as graph partitioning (e.g.~\cite{MVM11,LGT11,MOV12,Vis12}), maximum flows (e.g.~\cite{CKM+10,LRS13,Mad13}) and max cuts (e.g.~\cite{Tre08}). Its application in quantum computation, nevertheless, is still scarce. It would be exciting to see more quantum algorithms (especially exponentially faster ones) developed based on this elegant theory. 

\section*{Acknowledgments}

The author thanks Andrew Childs, David Gosset, Zeph Landau, Aaron Ostrander, Mario Szegedy and Umesh Vazirani for useful discussions and comments. The author is also grateful to Robin Kothari for suggesting the construction in the proof of Theorem \ref{thm:lowerbound}. Part of this work was done while the author was a graduate student at Computer Science Division, University of California, Berkeley. This research was supported by NSF Grant CCR-0905626 and ARO Grant W911NF- 09-1-0440.

\renewenvironment{thebibliography}[1]
        {\frenchspacing
         \small\rm\baselineskip=11pt
         \begin{list}{\arabic{enumi}.}
        {\usecounter{enumi}\setlength{\parsep}{0pt}     
         \setlength{\leftmargin}{17pt}  
                \setlength{\rightmargin}{0pt}
         \setlength{\itemsep}{0pt} \settowidth
          {\labelwidth}{#1.}\sloppy}}{\end{list}}
\nonumsection{References}
\bibliographystyle{unsrt}

\appendix
\noindent
The following lemma is used in Section \ref{sec:qlsa}. It says that if two unnormalized states are close and one of them has a large norm, then their normalized versions are also close.
\begin{lemma}
Let $\ket{\psi}$ and $\ket{\phi}$ be two unnormalized states satisfying $\nm{\ket{\psi}} \ge \alpha >0$ and $\nm{\ket{\psi}-\ket{\phi}} \le \beta$. Then 
\beba
\nm{\dfrac{\ket{\psi}}{\nm{\ket{\psi}}}-\dfrac{\ket{\phi}}{\nm{\ket{\phi}}}} \le \dfrac{2\beta}{\alpha}.
\eeea
\label{lem:statedistance}
\end{lemma}
\proof{ Using the triangle inequality, we get
\bea
\nm{\dfrac{\ket{\psi}}{\nm{\ket{\psi}}}-\dfrac{\ket{\phi}}{\nm{\ket{\phi}}}}
&=&
\nm{\dfrac{\ket{\psi}}{\nm{\ket{\psi}}}-\dfrac{\ket{\phi}}{\nm{\ket{\psi}}}+\dfrac{\ket{\phi}}{\nm{\ket{\psi}}}-
\dfrac{\ket{\phi}}{\nm{\ket{\phi}}}} \\
&\le &
\nm{\dfrac{\ket{\psi}}{\nm{\ket{\psi}}}-\dfrac{\ket{\phi}}{\nm{\ket{\psi}}}} + 
\nm{\dfrac{\ket{\phi}}{\nm{\ket{\psi}}}-
\dfrac{\ket{\phi}}{\nm{\ket{\phi}}}}  \\
& \le &
\dfrac{\nm{\ket{\psi}-\ket{\phi}}}{\nm{\ket{\psi}}} + 
\nm{\ket{\phi}}\abs{\dfrac{1}{\nm{\ket{\psi}}}-\dfrac{1}{\nm{\ket{\phi}}}}\\
&= &
\dfrac{\nm{\ket{\psi}-\ket{\phi}}}{\nm{\ket{\psi}}} + 
\dfrac{\abs{\nm{\ket{\psi}}-\nm{\ket{\phi}}}}{\nm{\ket{\psi}}}\\
&\le &
\dfrac{2\nm{\ket{\psi}-\ket{\phi}}}{\nm{\ket{\psi}}}\\
&=&
\dfrac{2\beta}{\alpha}.
\eea
}

\end{document}

%% file: projection_fig.tex
\usetikzlibrary{intersections,calc}

\newcommand{\RightAngle}[4][2pt]{%
        \draw ($#3!#1!#2$)
        --($ #3!2!($($#3!#1!#2$)!.5!($#3!#1!#4$)$) $)
        --($#3!#1!#4$) ;
        }

\tdplotsetmaincoords{70}{90}

\resizebox{0.85\textwidth}{!}{

\begin{tikzpicture}[scale=3,tdplot_main_coords,>=latex, x={(1,-0.5,0)},rotate around z = -15, rotate around x = 5]
    
    \fill[fill opacity=0.65, fill=cyan!60]
    (0.75,1.5,1.5)--(0.75,0,0)--(-0.75,0,0)--(-0.75,1.5,1.5)node[ above] {$L_2$}--cycle;
     \fill[fill opacity=0.65, fill=cyan!60]
    (0.75,0,0)--(0.75,-1,-1)--(-0.75,-1,-1)-- (-0.75,-0.535,-0.535)--cycle; 
    
    \fill[fill opacity=0.65, fill=blue!50]
    (0.75,3,0)-- (0.75,0,0) -- (-0.75,0,0)--(-0.75,3,0)node[above] {$L_1$}--cycle;
     \fill[fill opacity=0.65, fill=blue!50]
    (0.75,-1.2,0) -- (0.75,0,0)--(-0.75,-0.85,0)--(-0.75,-1.2,0) --cycle;
     
     
     
     
     \draw[color = blue!100 ] (0.75,3,0)--(0.75,-1.2,0); 
     \draw[color = blue!100 ] (0.75,3,0)--(-0.75,3,0); 
     \draw[color = blue!100 ] (-0.75,3,0)--(-0.75,0,0);
     \draw[color = blue!100 ] (-0.75,-0.85,0)--(-0.75,-1.2,0); 
     \draw[color = blue!100 ] (0.75,-1.2,0)--(-0.75,-1.2,0); 
      
      \draw[color = cyan!120 ] (0.75,1.5,1.5)--(0.75,-1,-1); 
      \draw[color = cyan!120  ] (0.75,-1,-1)--(-0.75,-1,-1); 
      \draw[color = cyan!120  ] (-0.75,-1,-1) --(-0.75,-0.535,-0.535);
      \draw[color = cyan!120  ] (-0.75,0,0)--(-0.75,1.5,1.5); 
      \draw[color = cyan!120  ] (-0.75,1.5,1.5)--(0.75,1.5,1.5); 
      
       \draw[ultra thick, red](-0.75,0,0)--(0.75,0,0);

    \filldraw[ultra thick] (0,0,0) circle (0.5pt) ++ (0,0.1,0.1) node[anchor=south east ]{$Y$};
    \filldraw[ultra thick] (0,0.6,0.6) circle (0.5pt) ++ (0,0.1,0.1) node{$X$};
    \filldraw[ultra thick] (0,0.55,0) circle (0.5pt) ++ (0,0.1,0.1) node[below left = 0.2cm]{$Z$};
    \filldraw[ultra thick] (0,1.45,0) circle (0.5pt) ++ (0,0.1,0.1) node{$O$};
    
     \draw[thick] (0,0.6,0.6)--(0,1.45,0);               
     \draw[thick,dashed] (0,0.6,0.6)--(0,0,0);       
     \draw[thick, dashed] (0,0.55,0)--(0,0,0);       
     \draw[thick, dashed] (0,0.55,0)--(0,1.45,0);  
     \draw[thick,dashed] (0,0.6,0.6)--(0,0.55,0);  

     \RightAngle{(0,1.45,0)} {(0,0.6,0.6)} {(0,0,0)};
     \RightAngle{(0,1.45,0)} {(0,0.55,0)} {(0,0.6,0.6)};
    
\end{tikzpicture}
}

%% file: lower_bound_fig.tex
\usetikzlibrary{positioning}
\tikzset{main node/.style={circle,fill=blue!25,draw,minimum size=1.2cm,inner sep=0pt, font = \large},
            }
 \tikzset{mid node/.style={circle,fill=green!60,draw,minimum size=1.2cm,inner sep=0pt, font = \large},
            }
 \tikzset{special node/.style={circle,fill=red!50,draw,minimum size=1.2cm,inner sep=0pt,font = \large},
            }
  \tikzset{empty node/.style={opacity =0,fill=white!100,draw,minimum size=0cm,inner sep=0pt},
            }

  \resizebox{0.85\textwidth}{!}{ 
            
  \begin{tikzpicture}
    \node[main node] (A1) { $(1^*,0)$};
    \node[main node] (A2) [right = 1cm  of A1]  { $(2^*,0)$};
    \node[main node] (A3) [right = 1cm  of A2]  {$(3^*,0)$};
    \node[main node] (A4) [right = 1cm  of A3]  {$(4^*,0)$};
    \node[main node] (A16) [right = 6cm  of A4]  {$(16^*,0)$};
    \node[main node] (A17) [right = 1cm  of A16]  {$(17^*,0)$};
    \node[main node] (A18) [right =1 cm of A17] {$(18^*,0)$};
    \node[main node] (A19) [right = 1cm of A18] {$(19^*,0)$};
    
    \node[special node] (B1) [below right = 2.3cm and 1.5cm of A1]  {$(1,0)$};
    \node[mid node] (B2) [right= 2cm  of B1] {$(2,0)$};
    \node[mid node] (B3) [right= 2cm  of B2] {$(3,0)$};
    \node[mid node] (B4) [right= 2cm  of B3] {$(4,0)$};
    \node[mid node] (B5) [right = 2cm  of B4] {$(5,0)$};
    \node[special node] (B6) [right = 2cm  of B5] {$(6,0)$};
    
    \node[mid  node] (C1) [below = 1.5cm  of B1]  {$(1,1)$};
    \node[mid node] (C2) [right= 2cm  of C1] {$(2,1)$};
    \node[mid node] (C3) [right= 2cm  of C2] {$(3,1)$};
    \node[mid node] (C4) [right= 2cm  of C3] {$(4,1)$};
    \node[mid node] (C5) [right = 2cm  of C4] {$(5,1)$};
    \node[mid node] (C6) [right = 2cm  of C5] {$(6,1)$};
    
    \node[main  node] (D1) [below left = 2.3cm and 1.5cm of C1]  {$(1^*,1)$};
    \node[main node] (D2) [right = 1cm  of D1]  {$(2^*,1)$};
    \node[main node] (D3) [right = 1cm  of D2]  {$(3^*,1)$};
    \node[main node] (D4) [right = 1cm  of D3]  {$(4^*,1)$};
    \node[main node] (D16) [right = 6cm  of D4]  {$(16^*,1)$};
    \node[main node] (D17) [right = 1cm  of D16]  {$(17^*,1)$};
    \node[main node] (D18) [right =1 cm of D17] {$(18^*,1)$};
    \node[main node] (D19) [right = 1cm of D18] {$(19^*,1)$};
    
     \node[empty node](A5)[right = 0.5cm  of A4]{};
     \node[empty node](A15)[right = 5cm  of A5]{};
     \node[empty node](D5)[right = 0.5cm  of D4]{};
     \node[empty node](D15)[right = 5cm  of D5]{};

    \path[draw]
    (A1) edge node {} (A2)
    (A2) edge node {} (A3)
    (A3) edge node {} (A4)
    (A16) edge node {} (A17)
    (A17) edge node {} (A18)
    (A18) edge node {} (A19)
    
    (D1) edge node {} (D2)
    (D2) edge node {} (D3)
    (D3) edge node {} (D4)
    (D16) edge node {} (D17)
    (D17) edge node {} (D18)
    (D18) edge node {} (D19)
    
    (B1) edge node {} (C2)
    (B2) edge node {} (C1)
    (B2) edge node {} (C3)
    (B3) edge node {} (C2)
    (B3) edge node {} (B4)
    (C3) edge node [below=10pt] {\Large $x_3 = 0$} (C4)
     
    (B4) edge node {} (C5)
    (B5) edge node {} (C4)
    
    (B5) edge node {} (B6)
    (C5) edge node [below=12pt] {\Large $x_5 = 0$} (C6);

    \path[draw, densely dashed]
    (A5) edge node {} (A15)
    (D5) edge node {} (D15);
    
    \path[draw, bend right =30]
     (A1) edge node {} (B1)
     (D19) edge node {} (C6);
     
     \path[draw, bend left =30]
     (D1) edge node {} (B1)
     (A19) edge node {} (B6);
     
     \path[draw opacity =0]
      (C1) edge node [below=12pt] {\Large $x_1 = 1$} (C2)
      (C2) edge node [below=12pt] {\Large $x_2 = 1$} (C3)
      (C4) edge node [below=12pt] {\Large $x_4 = 1$} (C5);
      
      \path[draw]
     (A4) edge node{} (A5)
     (A15) edge node{} (A16)
     (D4) edge node{} (D5)
     (D15) edge node{} (D16);

\end{tikzpicture}
}